\documentclass[twocolumn]{aastex63}
\accepted{November 24, 2020}
\received{September 15, 2020}
\revised{October 29, 2020}
\submitjournal{ApJS}

\shorttitle{}

\usepackage{color}		
\definecolor{midgray}{gray}{0.4}	
\definecolor{orange}{rgb}{1,0.5,0} 
\definecolor{blue}{rgb}{0,0,0.6}  
\usepackage{apjfonts}

\definecolor{ao}{rgb}{0.0, 0.0, 1.0}

\usepackage{scrextend}
\deffootnote[0.5em]{0em}{1em}{\textsuperscript{\thefootnotemark}\,}

\usepackage{etoolbox}
\makeatletter
\patchcmd{\NAT@citex}
  {\@citea\NAT@hyper@{\NAT@nmfmt{\NAT@nm}\NAT@date}}
  {\@citea\NAT@nmfmt{\NAT@nm}\NAT@hyper@{\NAT@date}}
  {}
  {}
\patchcmd{\NAT@citex}
  {\@citea\NAT@hyper@{%
     \NAT@nmfmt{\NAT@nm}%
     \hyper@natlinkbreak{\NAT@aysep\NAT@spacechar}{\@citeb\@extra@b@citeb}%
     \NAT@date}}
  {\@citea\NAT@nmfmt{\NAT@nm}%
   \NAT@aysep\NAT@spacechar%
   \NAT@hyper@{\NAT@date}}
  {}
  {}
\patchcmd{\NAT@citex}
  {\@citea\NAT@hyper@{%
     \NAT@nmfmt{\NAT@nm}%
     \hyper@natlinkbreak{\NAT@spacechar\NAT@@open\if*#1*\else#1\NAT@spacechar\fi}%
       {\@citeb\@extra@b@citeb}%
     \NAT@date}}
  {\@citea\NAT@nmfmt{\NAT@nm}%
   \NAT@spacechar\NAT@@open\if*#1*\else#1\NAT@spacechar\fi%
   \NAT@hyper@{\NAT@date}}
  {}
  {}
\makeatother

\makeatletter
\def\blfootnote{\xdef\@thefnmark{}\@footnotetext}
\makeatother

\newcommand{\myemail}{tmorishita@stsci.edu}
\newcommand{\simgt}{\,\rlap{\lower 3.5 pt \hbox{$\mathchar \sim$}} \raise
1pt \hbox {$>$}\,}
\newcommand{\simlt}{\,\rlap{\lower 3.5 pt \hbox{$\mathchar \sim$}} \raise
1pt \hbox {$<$}\,}
\newcommand{\Msun}{M_{\odot}}

\newcommand{\logm}{\log M_*/\Msun}

\newcommand{\spbg}{{SuperBoRG}}
\newcommand{\ly}{${\rm Ly\alpha}$}

\def\Nfld{316} 
\def\numfld{\Nfld} 
\def\Afld{$0.41\,\mathrm{deg}^2$} 
\def\Nspec{98} 
\def\Nobj{463662} 

\def\jh{$JH_{\rm 140}$} 
\def\hh{$H_{\rm 160}$} 

\newcommand{\hst}{{HST}}

\newcommand{\spit}{{Spitzer}}

\newcommand{\Z}{{\rm [Z/H]}}

\newcommand{\sext}{{SExtractor}}

\newcommand{\eazy}{{\ttfamily EAzY}}

\begin{document}
\title{ \Large
SuperBoRG: Search for The Brightest of Reionizing Galaxies and Quasars in HST Parallel Imaging Data$^{\dagger}$
}

\newcommand{\affilA}{Space Telescope Science Institute, 3700 San Martin Drive, Baltimore, MD 21218, USA; \href{mailto:\myemail}{\myemail}}

\author{T.~Morishita}
\affiliation{\rm \affilA}

\begin{abstract}
The {Hubble Space Telescope} (\hst) has been providing tremendous survey efficiency via its pure-parallel mode, by observing another field in parallel with the primary instrument in operation for the primary observation. In this study, we present a new archival project, \spbg, which aims at compiling data taken in extragalactic parallel programs of \hst\ with {Wide Field Camera 3} in the past decade; including {\it pure-parallel} (BoRG, HIPPIES, and COS-GTO) and {\it coordinated-parallel} (CLASH and RELICS) programs. The total effective area reaches $\sim$\,\Afld\ from  $4.1$\,Msec, or 47\,days, of observing time, which is the largest collection of optical-to-near-infrared imaging data of \hst\ for extragalactic science. We reduce all data in a consistent manner with an updated version of our data reduction pipeline, including a new sky background subtraction step. When available, imaging data from the \spit\ Space Telescope are also included in photometric analyses. The dataset consists of \Nfld\ {\it independent} sightlines and is highly effective for identification of high-$z$ luminous sources ($M_{\rm UV}\simlt-20$\,mag) at $z\sim7$ to $12$, helping to minimize the effects of cosmic variance. As a demonstration, we present three new  $z\simgt7$ source candidates, including one luminous galaxy candidate at $z_{\rm phot}\sim10.4$ with $M_{\rm UV} \sim -21.9$\,mag; the best-fit spectral energy distribution implies a large amount of stellar mass ($\logm\sim10$) and moderate dust attenuation ($A_{V}\sim1.4$\,mag), though the possibility of it being a low-$z$ interloper cannot completely be rejected ($\sim23\%$) with the current dataset. The dataset presented in this study is also suited for intermediate and low-$z$ science cases.
\end{abstract}

\section{Introduction}\label{sec:intro}
\blfootnote{$\dagger$ Based on observations made with the NASA/ESA Hubble Space Telescope, obtained from the data archive at the Space Telescope Science Institute. STScI is operated by the Association of Universities for Research in Astronomy, Inc. under NASA contract NAS 5-26555 {(doi: 10.17909/t9-m7tx-qb86)}.}

The formation of the first stars, black holes, and galaxies is one of the central questions in current astronomical research. In particular, our understanding has been significantly advanced from the discovery of galaxies and quasars \citep[e.g.,][]{oesch16,banados18,hashimoto18,yang20} in the epoch of reionization \citep[][]{gunn65,madau99,becker01,robertson15}.

Significant contribution to our exploration into such an early epoch has been made by the {Hubble Space Telescope} (\hst) since the installation of Wide Field Camera 3 (WFC3). The new camera has advanced our understanding of galaxies beyond the previous redshift limit, $z\sim7$-8, and provided glimpses of galaxy formation in the early universe. A tremendous investment with the instrument has revealed hundreds of galaxy candidates at $z\simgt7$ from various surveys; wide surveys like the Cosmic Assembly Near-infrared Deep Extragalactic Legacy Survey \citep[CANDELS;][]{koekemoer11,grogin11} and the Brightest of Reionizing Galaxies survey \citep[BoRG;][]{trenti11,bradley12}, and deep surveys the Hubble Ultra Deep Field 2012 \citep[HUDF12;][]{ellis13}, the Hubble eXtreme Deep Field \citep[XDF][]{illingworth13}, Cluster Lensing And Supernova survey with Hubble \citep[CLASH;][]{postman12}, the Hubble Frontier Fields \citep[HFF;][]{lotz17}, and Reionization Lensing Cluster Survey \citep[RELICS;][]{coe19}. Follow-up spectroscopic campaigns have successfully confirmed $\sim20$ of those candidates at $z>7$ \citep[e.g.,][]{ono12,finkelstein13,oesch15,stark15,oesch16,song16,hoag18,hashimoto18,tamura18}. Some of them sources are surprisingly luminous \citep{zitrin15,oesch16}, exhibiting new interesting aspects of galaxy formation in the early universe, in conjunction with the formation of supermassive black holes as well \citep{mortlock11,banados18,yang20}.

Bright sources are of particular interest from many perspectives; not only for their formation mechanisms in such an early epoch \citep[e.g.,][]{roberts-borsani20}, but for their inferential application to cosmic reionization \citep{treu13,konno14,mason18,hoag19,banados18,davies18}. While low escape fraction of \ly\ photons is often the case for typical galaxies of luminosity $L<L^*$, situations seem dramatically different for luminous sources; theoretically, luminous sources are able to create a large ionizing bubble, where a higher fraction of \ly\ photons can escape \citep{cen00,yajima18,mason20}. Indeed, \citet{stark17} found a significantly high fraction of \ly\ detection in galaxies at $M_{\rm UV}<-20.3$\,mag. A high escape fraction of luminous sources is also suggested from a double-peaked \ly\ line profile \citep[e.g.,][]{matthee18}. Therefore, luminous objects are ideal targets where we are likely to see \ly\ emission and gain insights into their properties \citep{mason18b}. 

Such luminous sources are, on the other hand, rare \citep[$\ll 10^{-5}$Mpc$^{-3}$;][]{bouwens15}, and thus significantly affected by cosmic variance \citep[][]{trenti08,robertson10,bhowmick20}. To obtain unbiased results from a single sightline, at least a $\sim1$\,deg$^2$ coverage will be needed at this redshift range \citep[see Fig.~3 of][]{robertson10}, which is prohibitively expensive with WFC3 with its FoV of $\sim4.7$\,arcmin$^2$. A good example for such significant cosmic variance is that \citet{roberts-borsani16} found three high-$z$ galaxies in one of the five CANDELS fields, EGS, whereas only one from the other four fields. \citet[][]{tilvi20} recently revealed an over density of three galaxies at $z=7.7$, again in the EGS field. These examples clearly demonstrate that a survey of many sightlines, as a supplement to legacy-type surveys with a small number of sightlines, is necessary for unbiased measurements.

\hst\ offers such an ideal opportunity via its pure-parallel observing mode---while the primary instrument is in operation, the secondary instrument can be used in parallel to observe a field a few arc-minutes away from the primary field. Since the coordinates of the pure-parallel field cannot be specified, pure-parallel opportunities are often used for identifying objects without any prior knowledge. That is, such opportunities are ideal to search for luminous rare objects in the early universe by minimizing the effects of cosmic variance \citep[e.g.,][]{atek11}. Previous pure-parallel observations in the BoRG program, which consist of $>1000$\,\hst\ orbits, have successfully identified $z>8$ galaxy candidates at the bright-end magnitude range, $M_{\rm UV}\sim-21$ to $-24$\,mag \citep{schmidt14,calvi16,bernard16,morishita18b}. Identification of such luminous sources from sufficiently large volume is critical to determine the shape of the luminosity function and its evolution \citep[][]{bowler14,bowler20,bouwens15,ren20}.

The latest study of the BoRG collaboration has further extended their focus from galaxy candidates to unresolved point-source candidates in search of quasars and intense starburst galaxies at $z\simgt8$ \citep{morishita20}. Identification of new quasars at high redshifts is of extreme importance---for example, theoretical studies expect that the evolution of the quasar luminosity function behaves differently depending on the mode of black hole evolution \citep[][]{volonteri10}. In \citet{morishita20}, they successfully identifies three point sources that satisfy color selection criteria for $z\sim8$ objects. While spectroscopic confirmation is not yet available, the work demonstrated a new angle of using pure-parallel data of \hst.

With its uniqueness and probing ability, pure-parallel fields with multi-band filters are of particular interest in the extragalactic community. Indeed, several follow-up studies have utilized data taken in previous BoRG campaigns and presented high-$z$ candidates independently selected by their criteria \citep[e.g.,][]{Rojas-Ruiz20}, and some with addition of new data \citep{livermore18,bridge19}. However, due to a number of difficulties in selecting high-$z$ source candidates, different studies ended up having different sets of source candidates, despite using the same dataset. Such inconsistencies are often attributed to differences in error analyses, source detection and flux measurement, color selection criteria, and photometric redshift codes. Even a small discrepancy could eventually lead to conflicting conclusions on, e.g., the shape of the luminosity function at the bright-end magnitude range because not a large number of samples is currently available.

To alleviate possible systematics, in this study we initiate a new archival project, \spbg. The primary purpose is to compile multi-band imaging data previously taken in {parallel} observations for extragalactic science cases, reduce them in a consistent manner, and publish data products for immediate use by the community. From both {\it pure-parallel} and {\it coordinated-parallels} programs with WFC3 in the past decade, we collect \numfld\,sightlines with multi-band imaging data with moderate depth, which becomes one of the largest area extragalactic surveys conducted with \hst. This unique dataset will be useful in many science cases; not only for high-$z$ source identification as previously presented by the BoRG collaboration, but for cosmology from clustering analysis \citep{robertson10,cameron19}, low- and intermediate-redshift galaxies, and even for foreground stars in our Galaxy \citep{ryan11,holwerda14b,vanvledder16}.

The contents of this paper are as follows: In Sec.~\ref{sec:data}, we present details on the data included in \spbg, with a short description for each of parallel programs, as well as follow-up data by \spit. In Sec.~\ref{sec:reduction}, we describe details on reduction processes, including several changes to our customized reduction pipeline since \citet{morishita18b}, and present improvements by comparing with previous data products. In Sec.~\ref{sec:photom}, we present photometric analysis, including photometry, redshift analysis, \spit\ photometry, which requires a separate reduction process due to its significantly large PSFs, and star-galaxy classification. In Sec.~\ref{sec:analysis}, as a demonstration of use of the data, we present new high-$z$ source candidates, and discuss possible caveats, while the entire list of high-$z$ candidates will be presented in a forthcoming paper. We describe the data products in Sec.~\ref{sec:products}, and summarize the paper in Sec.~\ref{sec:sum}. Throughout, we quote magnitudes in the AB system.

\begin{figure*}
\centering
	\includegraphics[width=0.8\textwidth]{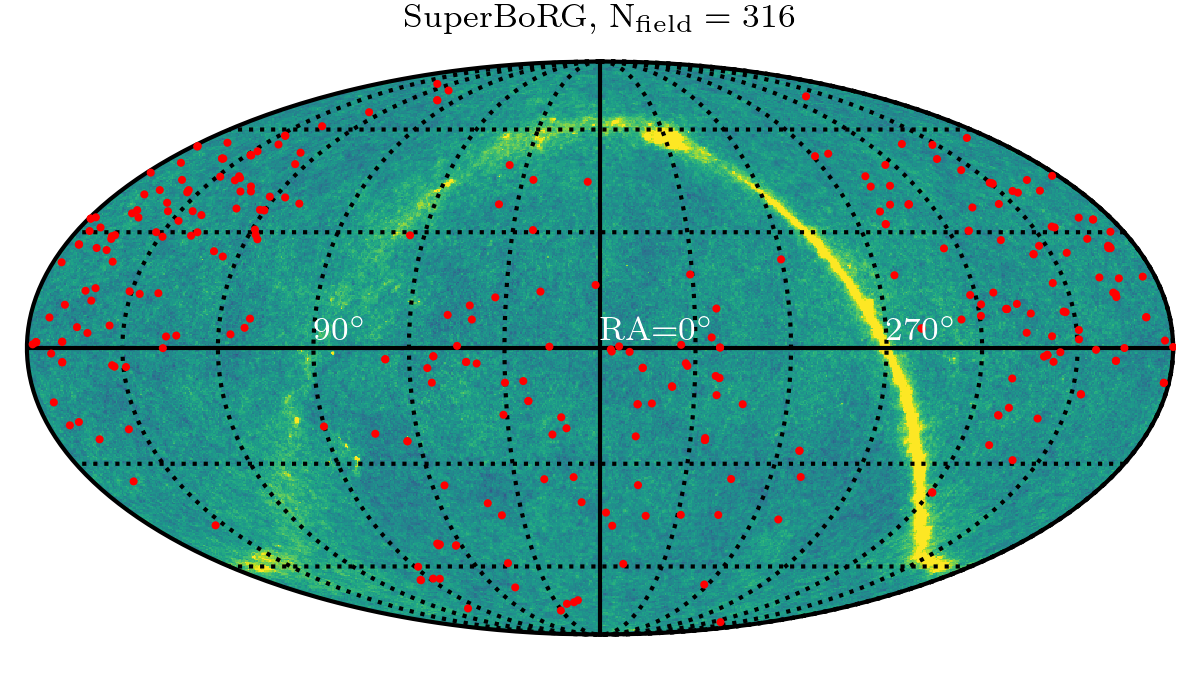}
	\caption{
	Distribution of \spbg\ fields (red circles) in the ecliptic coordinates, overlaid on a temperature map from the WMAP 5-year data \citep{hinshaw09}. Most of the fields were selected from high Galactic latitudes ($|b| > 30^\circ$), to avoid fields with significant Galactic extinction and/or dominated by foreground stars.}
\label{fig:field}
\end{figure*}

\begin{figure*}
\centering
	\includegraphics[width=0.41\textwidth]{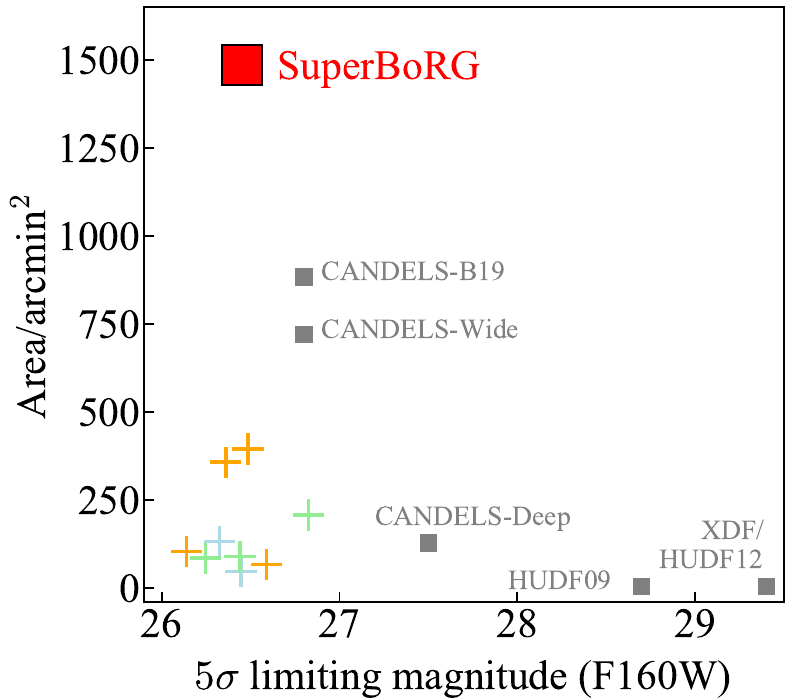}
	\includegraphics[width=0.49\textwidth]{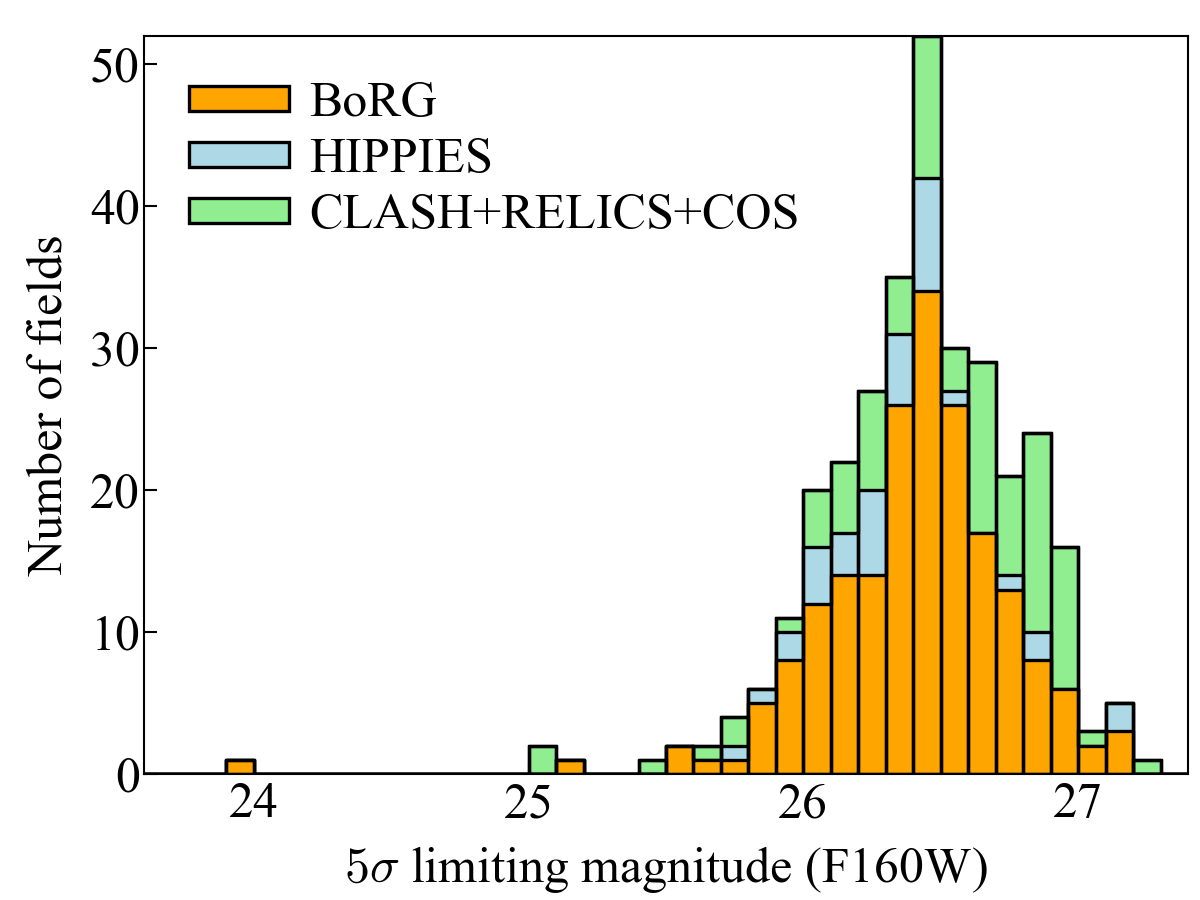}
	\caption{
	(Left): Effective area and $5\,\sigma$ limiting magnitude of various extragalactic survey programs of \hst. Survey programs included in \spbg\ (red square) are shown with cross symbols; BoRG cycles 17, 19, 22, 25 (orange), HIPPIES cycles 17, 18 (blue), CLASH, RELICS, and COS-GTO (green).  
	Legacy surveys such as CANDELS, including its extended version presented in \citet[][CANDELS-B19]{bouwens19}, and HUDF09/12/XDF programs are shown with gray squares.
	(Right): Histogram of limiting magnitude of the \spbg\ fields (Sec.~\ref{ssec:rms}). Survey programs are distinguished by the same color scheme as in the left panel. The whole list for field coordinates and limiting magnitudes is summarized in Table~\ref{tab:field}.
	}
\label{fig:hist}
\end{figure*}

\section{Data: HST pure-parallel observations}
\label{sec:data}
In this \spbg\ project, we collect fields from both pure-parallel and coordinated-parallel programs. The positional distribution in the sky is shown in Fig.~\ref{fig:field}, and area-magnitude distributions in Fig.~\ref{fig:hist}. These selected fields have a sufficient number of broadband filters and depth for extragalactic science.

\subsection{HST pure-parallel imaging data}\label{sec:hstdata}
\subsubsection{BoRG}\label{ssec:data_borg}
BoRG is a series of pure-parallel imaging programs initiated in 2009 \citep{trenti11,trenti12,bradley12}. These programs were designed to search for luminous galaxies in the epoch of reionization, namely at $z\simgt7$; thus observations were designed to efficiently identify high-$z$ objects via the Lyman break technique \citep{steidel96}, at a medium depth exposure with WFC3-UVIS and IR filters. Since its initiation, multiple programs have been awarded (Cycles 17, 19, 22, and 25; PIDs 11700, 12752, 13767, 15212), and the total number of \hst\ orbits collected in the entire BoRG programs reaches $\sim1400$.

Each of the BoRG programs has a different combination of filters, and thus selection criteria for high-$z$ galaxy candidates are different; however, there are basically three redshift ranges of interest, $z\sim8$, 9, and 10, based on non-detection in WFC3-UVIS filters at $\simlt1.0\,\mu{\rm m}$, and a color break measured with WFC3-IR filters. As an example, we present a few high-$z$ source candidates selected in this way in Sec.~\ref{ssec:cs}. 

Filter sets of BoRG are also suited for identifying $z\sim2$ galaxies, especially passively evolving galaxies with a strong Balmer break \citep[e.g.,][]{cameron19}. While the color break mimics the Lyman break at higher redshift, such degeneracy can be resolved by combining \spit\ data when available (Sec.~\ref{ssec:spit}).

The BoRG team has been adopting a carefully designed phase-II strategy since its initiation. For example, IR persistence in detectors is of particular concern because of the possibility of introducing an artificial coherent signal into the near-IR bands. Therefore, in each visit, they arranged the sequence of WFC3-IR filters to minimize the impact of persistence for the primary science goal, detection of dropout sources in blue filters such as F350LP and F105W. As detector persistence decays over time (with approximate power-law behavior), any saturated target observed in a previous visit most affects the initial part of the pure-parallel orbit. The general strategy therefore was to observe in these dropout filters as early as possible in each orbit \citep[see][for more details]{bradley12}.

It is noted that 7 fields of the Cycle 25 programs are discarded due to guide star acquisition failure during the entire visits. Many other fields were partially affected by similar failures. In fact, the fraction of guide star failures in the Cycle 25 fields is much higher than previous cycles, possibly due to the operational updates of \hst, which originated from reduction of the number of gyros since 2018. Therefore, some of the Cycle 25 fields do not reach the depth that were originally designed. 

Lastly, some of the BoRG fields had follow-up visits with WFC3 and ACS (PIDs 12905, 14652, 14701, 15702), primarily aiming at securing high-$z$ source candidates by adding medium-band imaging of WFC3-IR F098M \citep[][]{livermore18} and ACS F814W filter \citep[][]{bridge19}. We include these data sets and reduce along with the primary data set.

\subsubsection{HIPPIES}\label{ssec:data_hippies}
Hubble Infrared Pure Parallel Imaging Extragalactic Survey, HIPPIES, is also a series of pure-parallel imaging surveys carried out in Cycles 17 and 18 \citep[PIDs 11702, 12286][]{yan11,yan12}. While the filter sets of HIPPIES are similar to the one in the BoRG Cycle 17 campaign ($V$, $Y$, F125W, F160W), the program focused on fewer fields and spent longer exposure time in F098M (F105W) in Cycle 17 (Cycle 18), to secure $Y$-band dropouts. It is noted, due to a different phase-II design strategy, some of the data in Cycle 18 are significantly contaminated by IR persistence, and careful inspection is required when identifying dropout sources (See Sec.~\ref{ssec:cs}).

\subsubsection{COS-GTO program}\label{sssec:data_cosgto}
A part of the COS-GTO programs contains parallel imaging observations with WFC3 (PIDs 11519, 11520, 11524, 11528, 11530, 11533, 11534, 11541, 12036, 12024, 12025). The phase II of these programs was carefully designed for extragalactic science, in a similar way as in the BoRG programs (Trenti, M., private communication). The program provides 22\,fields with multi-band images at medium depth. Some of the fields from this program also include additional UVIS filters that are not frequently used in other programs in \spbg, such as F300X and F475X (with an extremely wide wavelength coverage), which improve sampling in spectral energy distribution (SED).

\subsection{HST coordinated-parallel imaging data}\label{ssec:copar}
Besides pure-parallel programs described above, we also include datasets from two coordinated-parallel programs spanning multiple fields: CLASH \citep{postman12} and RELICS \citep{coe19}. While the primary targets of these programs are fields of massive clusters of galaxies at $z<1$, associated parallel fields are located $\sim6$\,arcmin away from the cluster center, and magnification effect is only a few percent if any.

\subsubsection{CLASH program}\label{sssec:data_clash}
Multi-band images from WFC3 were taken in 44\,parallel fields\footnote{The number of parallel fields is larger than the number of targeted clusters, because many of the clusters have two different position angles, which produced more than one parallel field per cluster.} during the primary imaging of 25\,massive clusters of galaxies in the program Cluster Lensing And Supernova Survey With Hubble, CLASH \citep[PID 12065;][]{postman12}. Most of the parallel fields consist of F350LP, F125W, and F160W filters, and primarily aimed at searching for high-$z$ SNe \citep[e.g.,][]{graur14,strolger15,riess18}. While the filter set technically allows $z\sim10$ source selection (Sec~\ref{ssec:cs}), the fraction of low-$z$ interlopers is significantly high due to having a single non-detection filter and one color. All fields, however, have at least partial \spit\ IRAC ch1 and 2 coverage pointed at the cluster center (and ch3 and 4 for some fields), which improves the photometric redshift quality (Sec.~\ref{ssec:photz}).

Spectroscopic redshifts are also available for several fields, taken at VLT with VIMOS \citep{biviano13,balestra16,annunziatella16,monna17}, FORS2 \citep{biviano13}, and MUSE \citep{karman15,karman17,grillo15,caminha16}, as well as with \hst\ grisms through the GLASS campaign \citep{schmidt14b,treu15,abramson20}. We retrieve publicly available redshift catalogs by the CLASH team, and cross-match with sources detected in this study. Since these spectroscopic campaigns were conducted primarily on the cluster center, the coverage in the parallel fields is partial; still, we find $90$\,objects with a good spectroscopic quality (see Sec.~\ref{ssec:photz}).

It is noted that we exclude 6 fields of CLASH which are presented in the Hubble Frontier Fields \citep{coe15,lotz17}, as these imaging data are much deeper than the exposure time of typical fields in this study, and sophisticated catalogs are already available from several dedicated studies \citep{castellano16,shipley18,bradac19}.

\subsubsection{RELICS program}\label{sssec:data_clash}
Reionization Lensing Cluster Survey, RELICS \citep[PID 14096;][]{coe19,salmon20}, is a \hst\ Treasury Program, targeting 41\,clusters at $0.18<z<0.97$ with the total of 188 orbits. Among 41 clusters, parallel fields were observed in 18\,fields where no previous ACS imaging was available, spending $3$\,orbits for each. The filters are set to those in the BoRG cycle 22 program, optimized for high-$z$ galaxy search, where F350LP is used for optical non-detection and four IR filters (F105W, F125W, F140W, and F160W) are used for dropout and color estimates. All parallel fields of RELICS have \spit\ IRAC ch1 and ch2 coverage, taken in part of the SRELICS program (Brada{\v{c}} et al., in preparation). Although images and photometric catalogs by the RELICS team are already available for these parallel fields, we start from the initial data reduction to have a consistent data set across the different survey programs presented in this study.

\subsection{Spitzer IRAC data}\label{ssec:spit}
While selection of high-$z$ source candidates requires multi-band images by \hst, it is ideal to have additional photometric constraints at rest-frame optical wavelength range to effectively exclude low-$z$ interlopers with a similar SED (e.g., dusty star forming galaxies at $z\sim2$; Sec.~\ref{ssec:photz}), as well as to better characterize their stellar properties. Mid-infrared coverage by \spit/IRAC plays such an important role by constraining SED at $>3\mu{\rm m}$ \citep[e.g.,][]{mclure11,bradac14,bouwens15}. 

There has been an effort by the BoRG team of following up fields with promising candidates, as well as by one of the most recent \spit\ DDT programs, COMPLETE2 (PID 14045; PI. Stefanon, M.). Typically, each of such fields has $\sim1$\,hr exposure with IRAC ch1 and ch2, which is sufficient for identification of low-$z$ interlopers down to $m_{160}\simlt 26$\,mag, due to their red $m_{160}-[3.6]$ color. A few fields also have ch3 and ch4 coverage, taken for other targets nearby where these \spbg\ fields are coincidently located within the FoV. For each field of \spbg, we check availability of IRAC data, and reduce along with the primary \hst\ dataset when available (Sec.~\ref{sssec:spitphot}).

\begin{figure*}
\centering
	\includegraphics[width=0.4\textwidth]{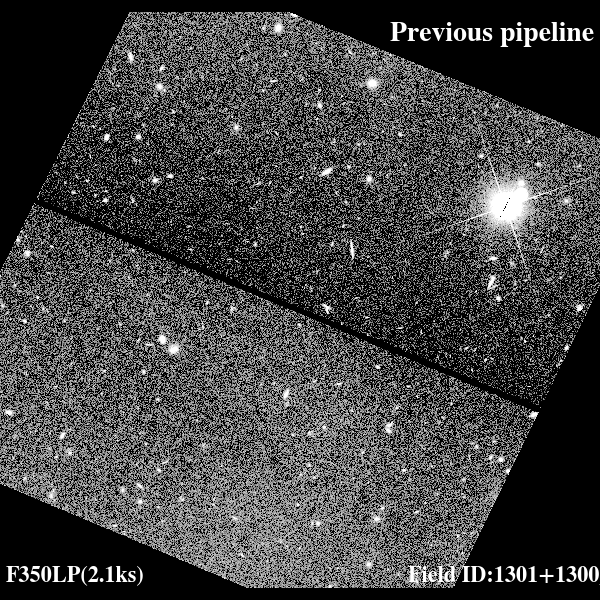}
	\includegraphics[width=0.4\textwidth]{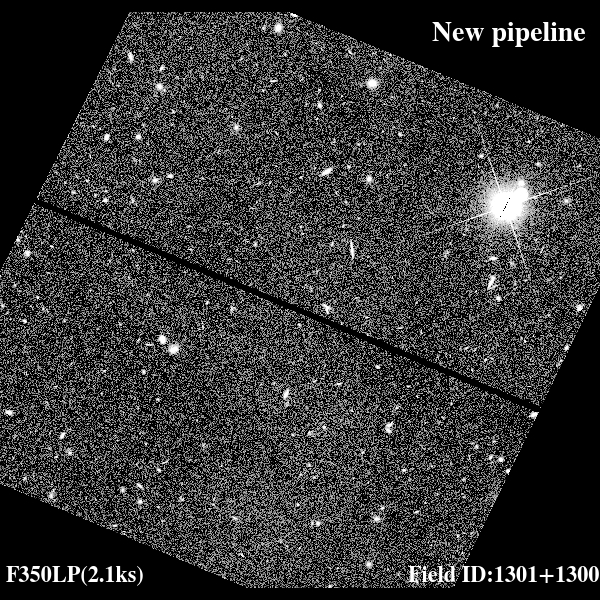}
	\includegraphics[width=0.4\textwidth]{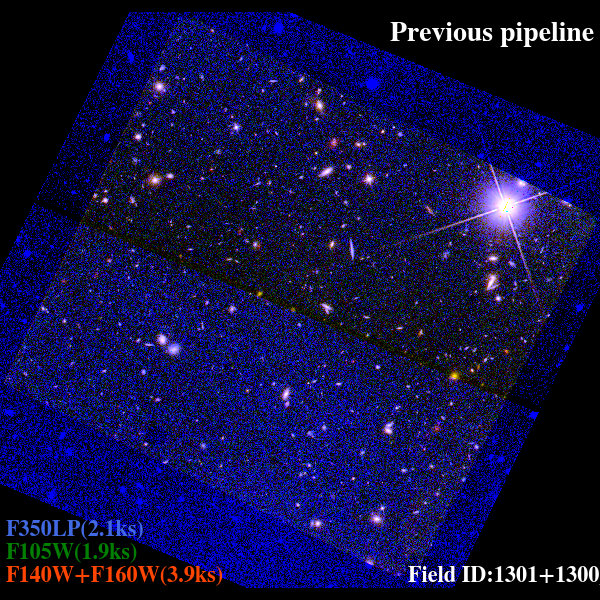}
	\includegraphics[width=0.4\textwidth]{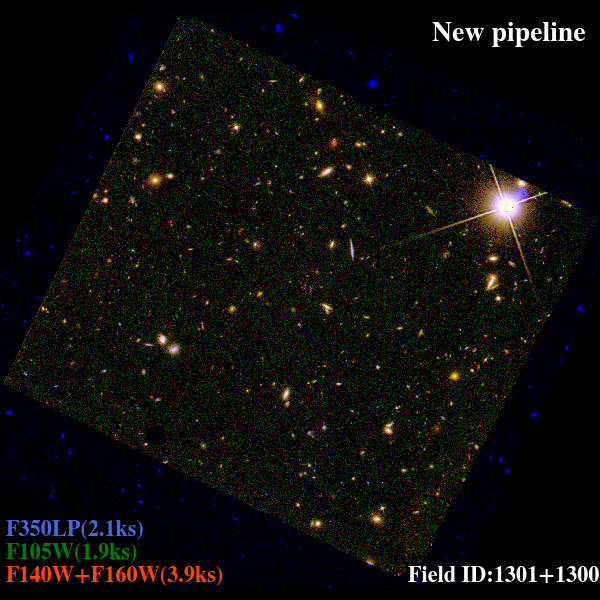}
	\caption{(Top) F350LP images of one of the \spbg\ fields, sBoRG-1301+1300. The image from the previous pipeline (left) shows spatially varying sky residual in images and mismatched sky residual level between the two UVIS chips. The new pipeline resolves the issues (right), which also improves image photometric quality (see Fig.~\ref{fig:bkg}). (Bottom) Pseudo color images of the same field. Non-zero sky residual in F350LP filter is seen in the previous pipeline product, which is resolved in the new pipeline product.}
\label{fig:rgb}
\end{figure*}

\begin{figure}
\centering
	\includegraphics[width=0.46\textwidth]{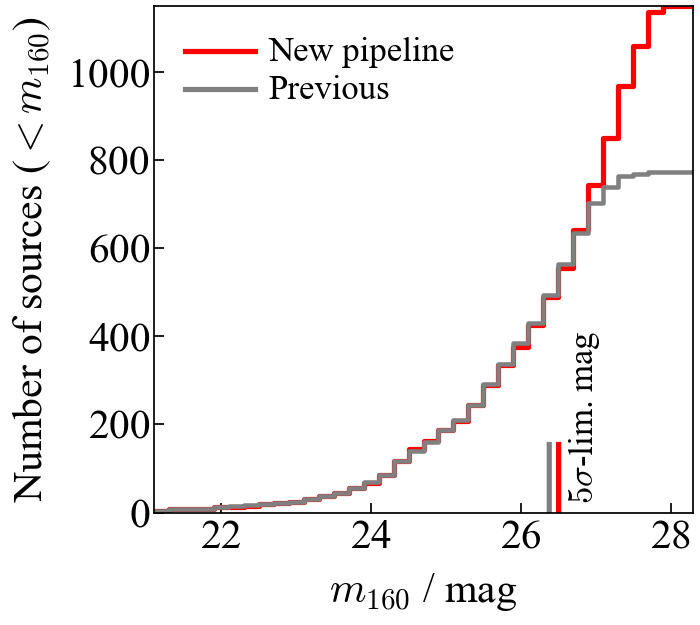}
	\caption{Cumulative distributions of sources detected in images from the previous pipeline (gray line) and the new pipeline (red) are shown. (The same field as in Fig.~\ref{fig:rgb}) The new pipeline product significantly improves detection at $\simgt27$\,mag, but also near the limiting magnitude, $\sim26.5$\,mag, which provides $\sim0.1$\,mag deeper limiting magnitude (vertical lines).}
\label{fig:bkg}
\end{figure}

\section{Reduction}\label{sec:reduction}
\subsection{HST image reduction}\label{ssec:imagereduction}
While pure-parallel opportunities provide us exceptional efficiency of mapping large areas of the sky, images taken in this mode are not dithered in most cases, as they cannot impact the primary observations, and thus careful post-processing is required. We use a dedicated  pipeline, {\tt Borgpipe}, originally designed for the BoRG data. A couple of updates are made in this study to improve the final data quality. The flow of the reduction process is as follows.

We start by visiting each field of the parallel programs above and retrieve all WFC3 and ACS images available via MAST. We download distortion-uncorrected imaging files: {\tt flt} for WFC3IR, and {\tt flc} for UVIS and ACS \citep[i.e. Charge-Transfer Efficiency-corrected images;][]{noeske12,anderson14}. We check two header flags, {\tt QUALITY} and {\tt EXPFLAG}, of each retrieved image to identify any issues during the exposure, such as guide star acquisition failure and unexpected interruption, and discard those with problems.

We then check any satellite trails crossing images by using an automated algorithm provided in ACStools.\footnote{\url{https://acstools.readthedocs.io/en/latest/}} 
The algorithm consists of three steps: 1.~edge detection by using a Canny algorithm \citep{canny86}, to find boundaries in any given image by detecting discontinuities in the intensities from pixel to pixel which define the edge of a feature in the image, 2.~clean step, by removing small objects whose perimeter is less than a certain number of connected pixels (75 by default), and 3.~a Probabilistic Hough Transform \citep{galamhos99}, to look for straight line segments in the image \citep[][for more details]{borncamp16}. We use a set of parameters suggested by the ACS team, which was found sufficiently effective in our initial test using a subset of data with satellite trails. Detected satellite trails in each flt/flc file are masked by changing the value in the DQ array so the contaminated pixels are not used in the following reduction. 

We then perform a customized cosmic ray (CR) rejection, by using the python version of the Laplacian edge filtering algorithm \citep[LACOSMIC;][]{vandokkum01}.\footnote{\url{http://lacosmic.readthedocs.io/en/latest/}} We set cr\_threshold to $2.3\,\sigma$ and $3.0\,\sigma$ for UVIS and IR detectors, respectively, from our dedicated test in \citet{morishita18b}. It is noted that an aggressive setup detection for the CR detection in the optical band may cause false identification of dropout sources by removing positive pixels in bluer bands as CRs.

These cleaned images are then aligned within each filter by using {\tt Tweakreg} and then combined. The combined images are then aligned to the F160W image before the final drizzle step. We also attempt to align the combined images to GAIA DR2 astrometry frame \citep{gaia18} when more than three stars are available. While image distortion and misalignment are automatically corrected by {\tt Tweakreg}, we carefully inspect the final data product, and tune configuration parameters field-by-field when necessary.

While {\tt astrodrizzle} performs sky background subtraction during the final drizzle step, it only subtracts a global sky value estimated from the entire detector. We found that some images show noticeable amounts of sky flux spatially varying across the detector (Fig.~\ref{fig:rgb}). Also, the two UVIS chips often show mismatched residual flux levels, which implies a more sophisticated method of sky background subtraction is required. To overcome this issue, we develop an alternative approach for sky subtraction by using \sext. For each flt/flc image we run \sext\ and subtract the output background image from the original image. Parameters of \sext\ for source detection and background size at this stage are very critical to final output products. We carefully accessed results in a subset of data, and optimized these parameters to {\tt BACK\_FILTERSIZE} 6 (10) and {\tt BACK\_SIZE} 250 (500) for WFC3-IR (WFC3-UVIS and ACS) data. The comparison between the original sky-subtraction method is shown in Sec.~\ref{sssec:bkg}.

Lastly, sky-subtracted images are combined with a common pixel scale of $0.\!\arcsec08$\,/\,pixel and with pixfrac to 0.75, optimized for our non-dithered images from previous studies. The sky subtraction step here is skipped to avoid duplication. This final step produces science and rms maps for each filter band, as well as a combined F140W+F160W image for source detection and optical-stacked image (combined with filters bluer than F098M) for optical non-detection of high-$z$ candidates.

\subsubsection{Background subtraction}\label{sssec:bkg}
In Fig.~\ref{fig:rgb}, we show one example field of \spbg\ that shows significant background residual in F350LP. The final image processed with the previous pipeline shows spatially varying background residual across each of the two UVIS detector chips. In addition, the residual levels of the two chips are not matched. These residual features are clearly noticeable in a pseudo rgb color image shown in Fig.~\ref{fig:rgb}. The final images processed with our custom sky subtraction step are shown in the right panels, where we can see uniform flux distribution across the detector. 

Figure~\ref{fig:bkg} shows magnitude distributions of sources detected in F160W images from the previous and new pipelines. While the number of detected sources in the new pipeline image remains similar at $\simlt 26$\,mag, significant increase is seen at fainter magnitude, by detecting additional $\simgt400$ sources in total. While these additional sources are fainter than the flux limit for science use in most cases, it is still important to locate these sources and, for example, mask when necessary to avoid any flux blending to a source of interest. An error analysis of the images (see Sec.~\ref{ssec:rms} for details) indeed reveals $\sim0.1$\,mag improvement in its limiting magnitude.

\begin{figure*}
\centering
	\includegraphics[width=0.98\textwidth]{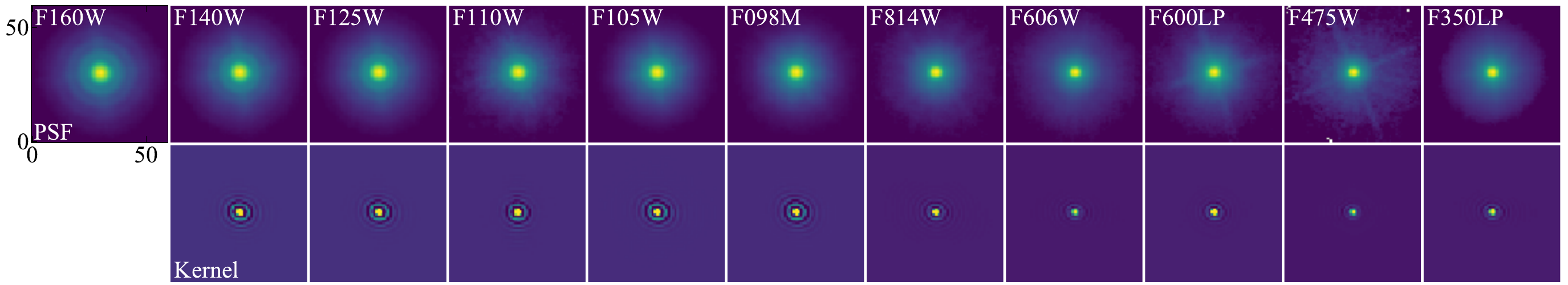}
	\caption{(Top) Median-combined PSFs of primary filters. Individual PSFs are collected from \spbg\ fields and then cross-matched with the GAIA DR2 catalog to ensure the selection of stars. (Bottom) Convolution kernels for each PSF used for PSF-matching to the F160W PSF, generated by {\ttfamily pypher}. 
	}
\label{fig:psfall}
\end{figure*}

\begin{figure}
\centering
	\includegraphics[width=0.48\textwidth]{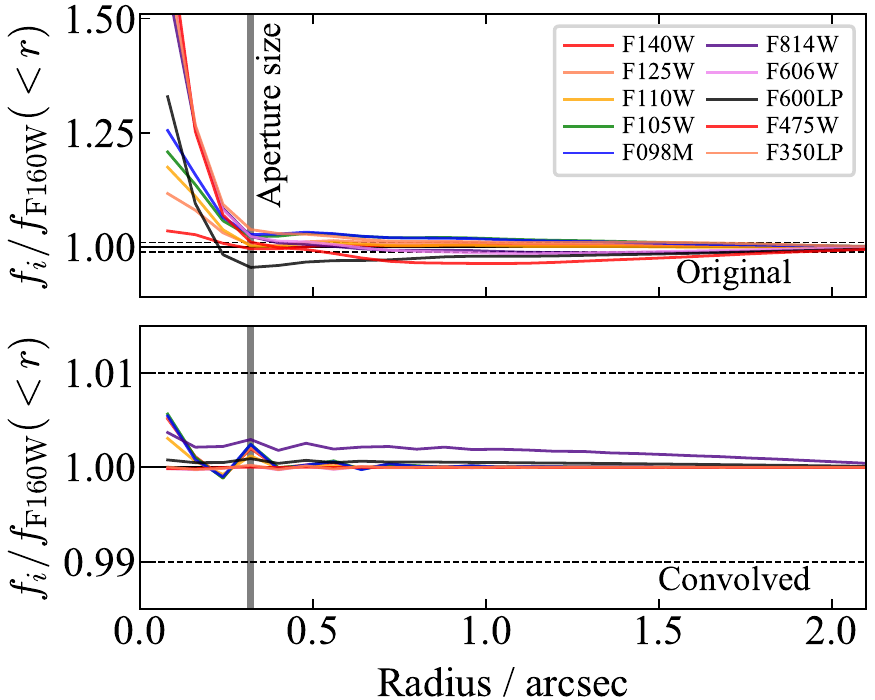}
	\caption{(Top) Total flux of original PSFs normalized by flux of the F160W PSF within a radius. (Bottom) Same as top panel, but after convolving PSFs with kernels generated by {\ttfamily pypher}. The difference remains $<0.3\%$ for all filters at a radius of aperture photometry, $r=0.\!\arcsec32$.
	}
\label{fig:psf}
\end{figure}

\subsubsection{PSF matching of images}\label{ssec:psf}
In the previous works of BoRG, images were not matched to a common point-spread function (PSF) size, partially due to lack of accurate PSF characterization. Instead, flux measurement has been conducted by the ISOPHOTAL mode of \sext, which collects flux in a segment defined in the detection image. However, flux estimated in ISOPHOTAL can be significantly underestimated in blue filters, which may cause an artificial dropout, especially when images are not dithered (Sec.~\ref{sssec:aper}). Since we collect a sufficiently large number of fields through a consistent reduction process, we here endeavor to improve photometric quality by matching image PSFs and adopting a canonical aperture photometry.

To have image PSFs matched between different filters, we first collect PSFs from the entire \spbg\ fields. We identify bright, but not saturated stars ($16.5<m/{\rm mag}<22$) from reduced images. To make sure only stars are collected, we cross-match identified objects with the GAIA DR2 catalog \citep{gaia18}. For each of those selected stars, we cut out into a postage stamp, subtract its local sky, and normalize the total flux. Due to a relatively large pixel size of \spbg, we resample each star into a $10\times$ smaller pixel scale size and shift the image to its light-weighted center before the final median-stack in each filter. The median-stacked PSFs of our primary filters are shown in Fig.~\ref{fig:psfall}. These PSFs are then provided to {\tt Pypher} \citep{boucaud16}, to generate matching kernels to the F160W PSF (also shown in Fig.~\ref{fig:psfall}). We investigate encircled fluxes of convolved PSFs in Figure~\ref{fig:psf}. The radial light profiles of convolved PSFs show a difference of $<0.1\%$ from the F160W PSF at radius $r=0.\!\arcsec32$ i.e. aperture size for our photometry (Sec.~\ref{ssec:photom}).

Due to a small number of available images, some filters do not have sufficient stars for characterizing its PSF. We therefore approximate these PSFs by using one close to these filters (F350LP for F300X and F435W, F606W for F555W and F625W, and F814W for F775W and F850LP). 
An alternative solution for this would be Tiny Tim PSFs \citep{krist95}. However, it is known that Tiny Tim PSFs show non-negligible offsets at the inner part of a light profile \citep{vanderwel12,bruce12,morishita14}. To avoid such additional uncertainty, we proceed with the empirical approximation for these filters.
It is noted that these filters appear only in a couple of fields. Furthermore, these filters are primarily used for non-detection in the selection of high-$z$ source candidates, and thus our approximations of these PSFs do not significantly affect the final result.

\subsubsection{Error analysis}\label{ssec:rms}
It is known that photometric flux error of \hst\ images measured by \sext\ is underestimated, primarily due to correlation between pixels \citep[][]{casertano00}. Root mean square (RMS) maps generated by our pipeline thus need to be scaled so that it represents more realistic uncertainty.

To estimate the scaling factor, we follow the method presented in \citet[][]{trenti11}. Briefly, we place 100\,apertures in randomly selected empty (i.e. devoid of real sources) regions of each image, and measure fluxes therein by using \sext\ \citep{bertin96}. The configuration parameters and aperture diameter size are set identical to those used for detection of real sources ($0.\!\arcsec64$; Sec.~\ref{ssec:photom}). Each RMS map is then scaled so that the standard deviation of fluxes of the empty apertures ($\langle f_{aper}\rangle$) and the median of flux errors returned by \sext\ ($\langle e_{aper}\rangle$) are matched. We repeat this analysis for 10\,times, with a new set of random apertures, and take the median of scaling factors calculated at each realization. Scaling factors here are generally larger than previous studies \citep[e.g.,][]{morishita18b}. We attribute this to the PSF matching process in this study, which increases correlation between pixels. The distribution of $e_{aper}$ is also used to estimate limiting magnitudes for the following analysis. A histogram of F160W limiting magnitudes ($5\,\sigma$) of the entire \spbg\ fields is shown in Fig.~\ref{fig:field}, characterizing $\sim26.5$\,mag as the median limiting magnitude.

\subsection{Spitzer image reduction}\label{ssec:spitred}
We check IRAC image availability for every \spbg\ field at the Spitzer heritage archive\footnote{\url{https://sha.ipac.caltech.edu/applications/Spitzer/SHA/}}, and retrieve imaging data of IRAC channels 1 and 2, in the level1 format (pbc) when available. We use the Spitzer Science Center reduction software MOPEX \citep{makovoz05} to reduce images. After background subtraction and alignment of each pbc image, we combine images by setting the final pixel scale to $0.\!\arcsec54$ and pixel fraction to 0.4. Combined images are then aligned to the detection image of \hst\ in the field. 

The combined images are then resampled to the same pixel size of \hst, $0.\!\arcsec08$, as a matched pixel scale is required by the flux extraction code for these images (Sec.~\ref{sssec:spitphot}). We use an astropy module, {\tt reproject}\footnote{\url{https://github.com/astropy/reproject}} for image resampling. As for \hst\ images, \spit\ images need correction for the correlated noise. RMS scaling factors and limiting magnitudes are estimated in the same manner as for \hst\ images, by measuring RMS in empty regions, but using a larger aperture of  diameter size $0.\!\arcsec54$ (Sec~\ref{ssec:rms}). 
Among the \spbg\ fields, 156 (121) fields have IRAC ch1 (ch2) coverage. Limiting magnitudes of IRAC images are summarized in Table~\ref{tab:field}. 
 
\section{Photometric properties}\label{sec:photom}

\subsection{HST Photometry}\label{ssec:photom}
For each field, sources are detected with \sext\ in the F140W+F160W detection image. Detection parameters are set as follows---{DETECT\_MINAREA}\,$=9$, {NTHRESH}\,$=0.7\,\sigma$, {DEBLEND\_NTHRESH}\,$=32$, {DEBLEND\_MINCONT}\,$=0.001$. A relatively large convolution size ($5$\,pixels for FWHM of Gaussian) for detection is used to reduce false detection of, e.g., discrete noise at the edge of the detector and residual cosmic rays. Photometry for each filter is conducted in the dual-imaging mode based on the detection image. Signal-to-noise ratios are calculated from flux and error measured in an aperture with diameter size of $0.\!\arcsec64$. We scale measured flux to the total flux, by multiplying the ratio of total flux ($f_{\rm auto,F160W}$) and aperture flux ($f_{\rm aper,F160W}$) measured in F160W band uniformly to all filters.

Once fluxes are measured and scaled, Galactic dust reddening is corrected by using the attenuation value retrieved for each coordinate from NED \citep{schlegel98,schlafly11}\footnote{\url{http://irsa.ipac.caltech.edu/applications/DUST/}}. We adopt the canonical Milky Way dust law \citep{cardelli89}.

\begin{figure*}
\centering
	\includegraphics[width=0.9\textwidth]{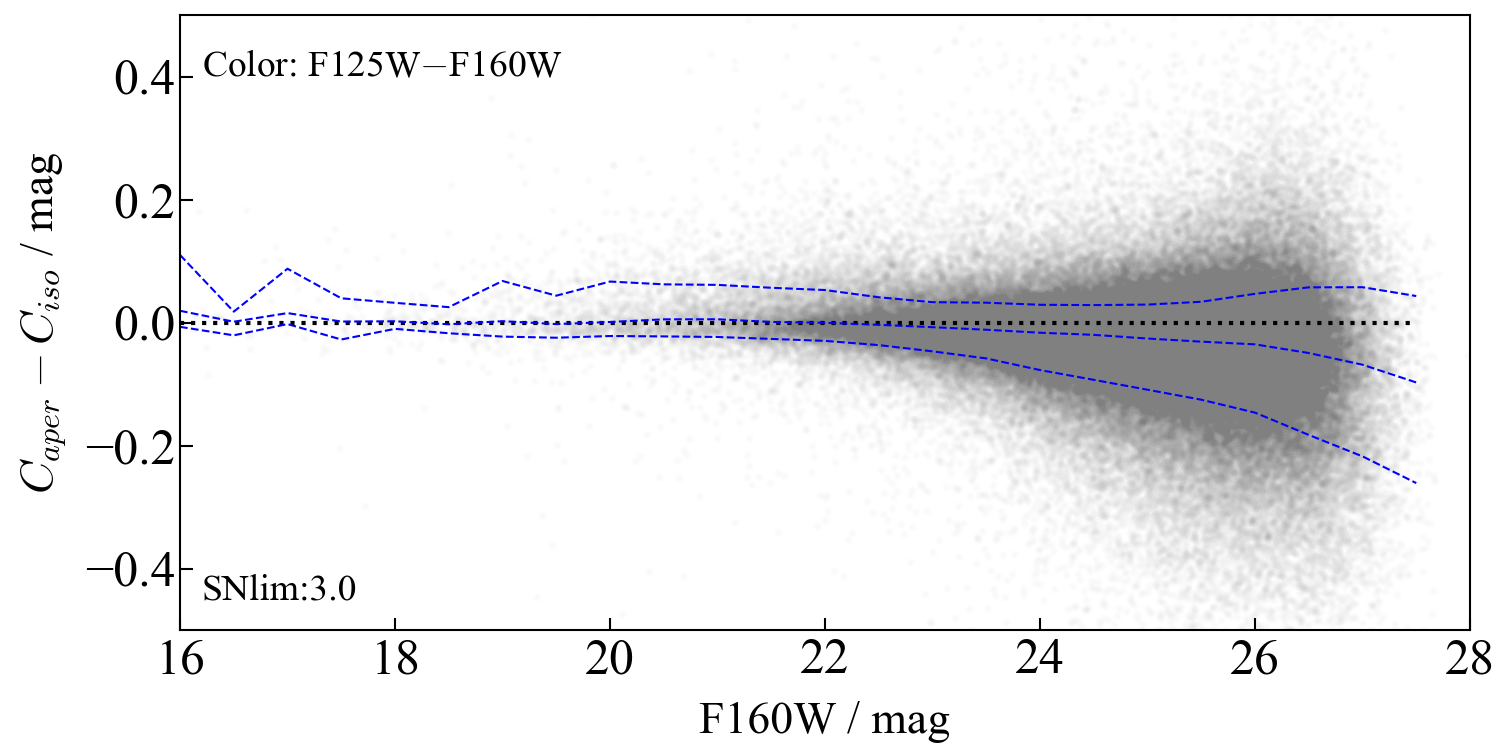}
	\caption{
	Comparison of aperture and isophotal photometry for sources with $S/N_{\rm F160W}>3$ and $\rm f_{persist}=0$. Colors, here F125W$-$F160W, are broadly consistent down to $\sim24$\,mag. At fainter magnitudes, however, the color measured by the isophotal photometry ($C_{\rm iso}$) systematically becomes redder than the one by the aperture photometry ($C_{\rm aper}$), which could result in false identification of high-$z$ dropout sources (Sec.~\ref{sssec:aper}). Running 16/50/84th percentiles (blue dashed lines) and the zero point (black dotted line) are shown.
	}
\label{fig:isoaper}
\end{figure*}

\begin{figure*}
\centering
	\includegraphics[width=0.9\textwidth]{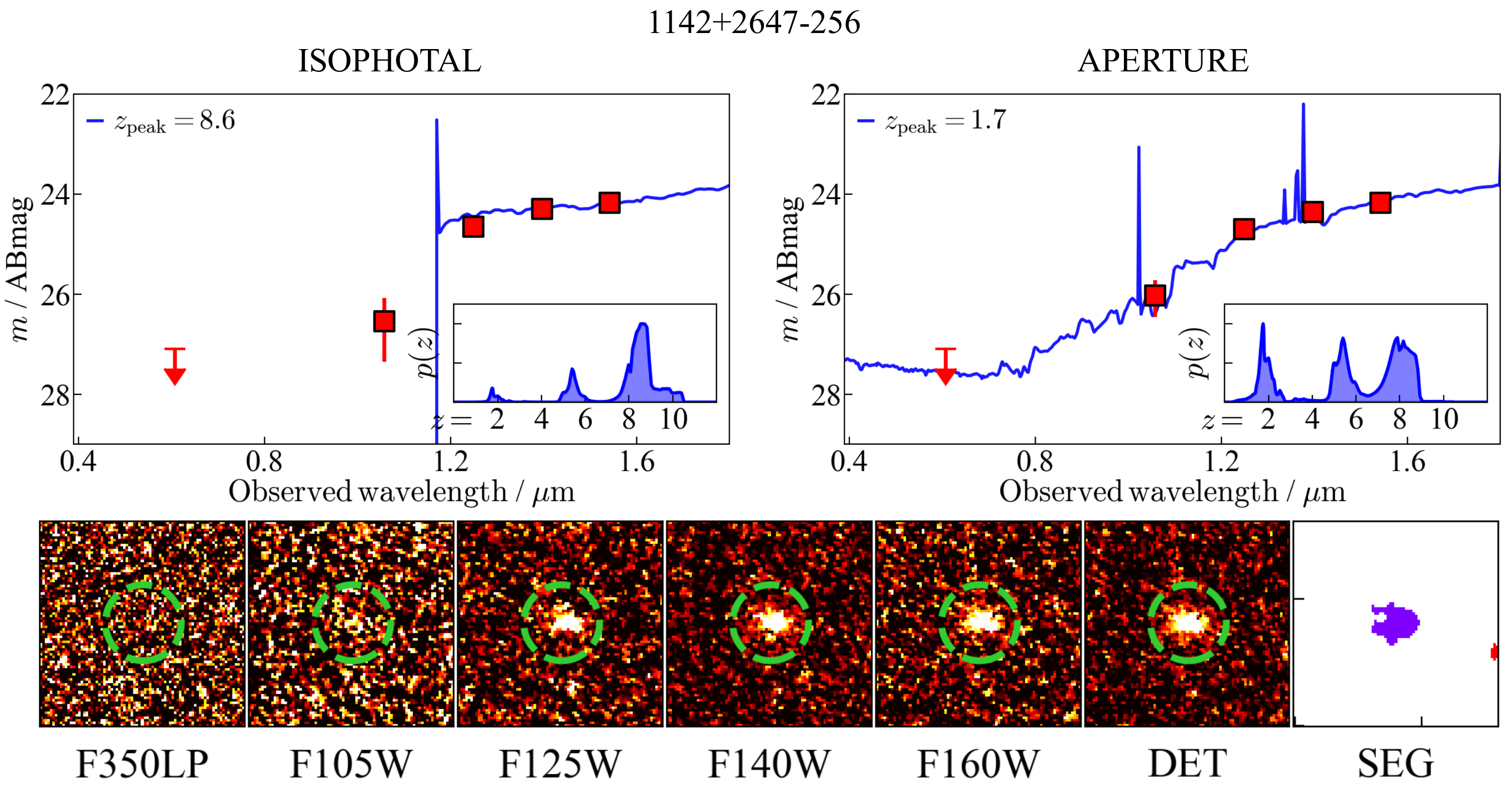}
	\caption{
	Example of misidentification of a low-$z$ source as a high-$z$ source candidate by adopting isophotal flux. This object, sBoRG-1142+2647-256, was originally identified as a z8\_Y105 candidate, by its strong dropout in F105W isophotal flux, with a primary peak at $z=8.6$ (left). The object, however, is excluded from the selection when aperture flux is used; F105W flux becomes more significant, and it is more likely a low-$z$ interloper, with a primary peak at $z=1.7$ (right). Such misidentification can happen because the isophotal measurement collects flux from pixels defined in the segmentation map (``SEG") uniformly for all filters, where more noise would be involved in bluer filers and lead to underestimation of S/Ns. On the other hand, aperture photometry collects flux from a defined aperture ($r=0.\!\arcsec32$; half the size of green circles), which is typically smaller than the area defined in the segmentation map, and provides more reliable flux measurements.
	}
\label{fig:flux}
\end{figure*}

\subsubsection{Comparison between aperture and isophotal fluxes}\label{sssec:aper}
In previous studies of BoRG, the isophotal photometry (FLUX\_ISO of \sext) has been a standard way of flux measuring. The isophotal photometry is known to accurately measure colors based on the detection map inferred from the detection image when the dual-imaging mode of \sext\ is used. In addition, it does not necessarily require images to be PSF-matched beforehand, whereas this is not the case for aperture photometry, where having a consistent central light profile is critical for unbiased measurements. 
While the isophotal photometry has often been proved effective when images are dithered, this is not always true for non-dithered images, as having a large detection area may introduce contamination by unmasked bad pixels and CRs.

To investigate this, we first compare colors (here F125W$-$F160W) derived with isophotal and aperture photometry in Fig.~\ref{fig:isoaper}. As is expected, these colors are broadly consistent down to $\sim24$\,mag. However, at a fainter magnitude range, which is typical for high-$z$ sources, the color derived by isophotal photometry is systematically redder than the aperture color, implying that the former could identify more high-$z$ source candidates.

In Fig.~\ref{fig:flux}, we showcase an example of false identification of a high-$z$ source candidate by this effect. The object was originally identified as a $z\sim8$ candidate, from its striking break color in F105W$-$F125W measured with the isophotal photometry. Indeed, photometric redshift analysis shows a primary peak at $z=8.6$. The object, however, is not selected as a high-$z$ candidate when the aperture photometry is used instead, as the estimated flux of F105W becomes more significant (i.e bluer break color), and it is found more likely to be a low-$z$ interloper at $z\sim2$.

Such misidentification of high-$z$ sources \citep[see also Appendix of][]{morishita18b} probably occurs because isophotal flux collects flux from pixels defined in a segmentation map (``SEG" in Fig.~\ref{fig:flux}), uniformly in all filters. Segmentation maps are often defined in red filters (here F140W+F160W image), and the region defined in such a way is larger than the actual size of the object in blue filters for many extragalactic sources. This would then include more empty pixels (i.e. noise) in flux measurement in bluer filers, and lead to underestimation of S/N. On the other hand, aperture photometry collects flux in an aperture defined, and by having a reasonably small aperture, such bias can be mitigated, though it requires images having matched PSFs. Given that our PSF-matching quality turns out to have $\ll1\%$ accuracy from the previous section, we choose aperture photometry for the following photometric analysis.

\subsection{Spitzer Photometry}\label{sssec:spitphot}
Since IRAC images have much larger PSFs ($0.\!\arcsec18$), deblending source flux from surrounding sources is a critical step for accurate characterization. To extract source flux in IRAC images, we use the T-PHOT package \citep{merlin16}. TPHOT estimates flux in a low-resolution (LR) image based on structure in a high-resolution (HR) image. We use the \jh+\hh\ detection image as the HR image for each field, and convolution kernels required by TPHOT are generated by stacking PSFs found in IRAC images. We resample \spit\ images in the same scale as for the HR image and provide those to TPHOT. 

We extract sources down to $m_{160}=24$\,mag uniformly for all fields where IRAC images are available. Those with extracted flux fainter than the limiting magnitude of each IRAC image, however, are replaced with the limiting magnitude as upper limit. TPHOT also returns a flag for extraction, to inform the user if the target object is blended with neighboring ones. We include the flag for all IRAC channels in the final photometric catalog.

For sources fainter than $m_{160}=24$\,mag, we do not perform the flux extraction process by TPHOT, as including such faint sources could occasionally cause a crash during the fitting stage. However, high-$z$ source candidates identified in \spbg\ (Sec.~\ref{ssec:cs}) are often fainter than this magnitude, and an extra flux extraction process is required. For those selected as high-$z$ source candidates, we follow a method presented by \citet{morishita20}, which models the LR light profile using GALFIT based on the profile parameters derived from HR images.\footnote{The code is in progress for public release and documentation.} Since this method concerns a smaller cutout and is computationally intensive, we only use this extraction method for sources of particular interest (Sec.~\ref{sssec:res}).

\begin{figure*}
\centering
	\includegraphics[width=0.9\textwidth]{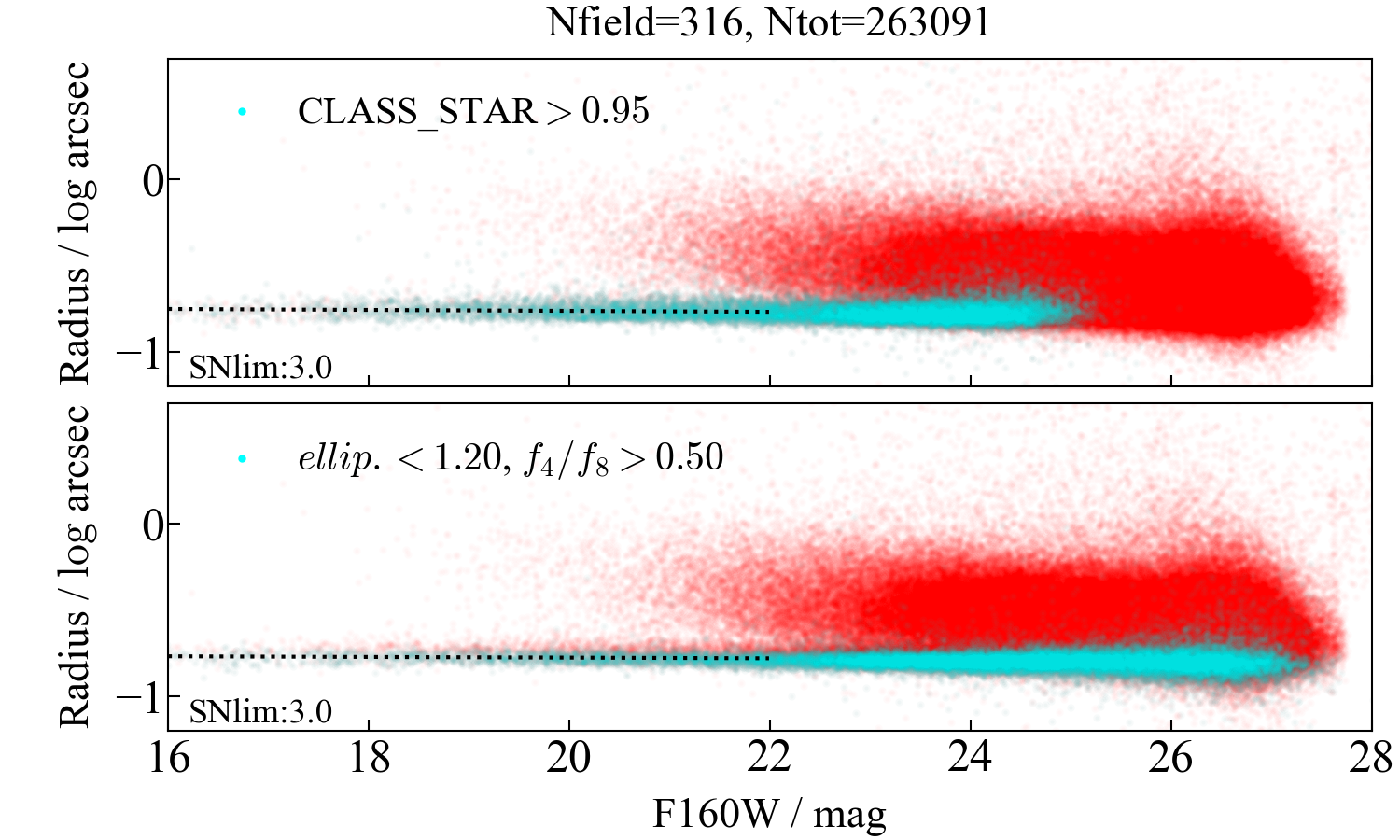}
	\caption{
	Distribution of sources with $S/N_{\rm F160W}>3$ and $\rm f_{persist}=0$ (red dots), in a size-magnitude plot. Those selected as point sources using CLASS\_STAR (top) and a combination of ellipticity and flux concentration (bottom) are highlighted (cyan dots). Slopes fit to the selected point sources are also shown (black dashed lines).
	}
\label{fig:size}
\end{figure*}

\subsection{{Point-source} selection}\label{ssec:size}
In Fig.~\ref{fig:size}, we show the size distribution of all sources with $S/N>3$, where the size is a non-parametric half-light radius (FLUX\_RADIUS) derived with \sext. Traditionally, CLASS\_STAR (a float number that ranges from 0 to 1, where a source is more likely a point source as the number increases) has been used as a useful parameter for star-galaxy classification. Those with CLASS\_STAR\,$>0.95$ tightly distribute at the lower edge of the distribution up to $\sim24$\,mag. However, as previously indicated \citep{finkelstein15,morishita20}, the flag is not complete at fainter magnitudes.

Instead, \citet{morishita20} demonstrate an alternative classification for point sources using ellipticity ($e$) and flux concentration, specifically,
\begin{equation}\label{eq:ps}
e<1.2 \wedge f_4/f_8>0.5
\end{equation}
where $f_{x}$ is flux measured within an aperture of $r=x$\,pixels. Sources selected with the criteria are shown in the bottom panel of Fig.~\ref{fig:size}, where we see a tight distribution extending down to a fainter magnitude, $\sim26$\,mag, providing a good alternative way of separating point and extended sources. {With this optimization down to fainter magnitudes, this criterion, in combination with color criteria presented below, can also be applied for high-$z$ point-source selection, in search of quasars and starburst galaxies \citep[see Section~\ref{ssec:cs}, and][for comprehensive analyses and discussion]{morishita20}.}

It is noted that there is a fraction of sources with CLASS\_STAR\,$>0.95$ that do not satisfy the criteria at the bright-end magnitude range, $<22$\,mag. Most of such sources are on the upper side of the sequence, which implies that the method offers a more conservative, but possibly incomplete way of separating point sources. It is noted that sources selected in both selection criteria show a flat slope of similar intercepts ($\log r / {\rm arcsec}= -0.76$ and $-0.78$, respectively). Sources with measured size below this limit should be considered as unresolved.


\begin{figure}
\centering
	\includegraphics[width=0.49\textwidth]{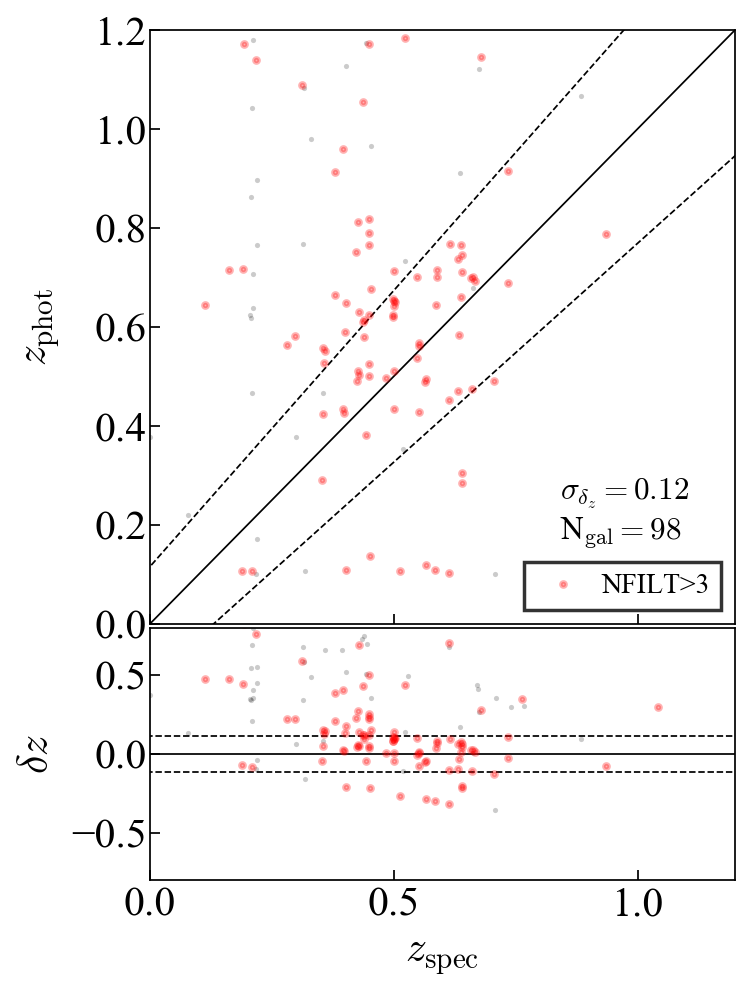}
	\caption{Comparison of photometric and spectroscopic redshifts of objects taken from the CLASH fields. For photometric redshift, we use  {\tt zmc}, which is calculated based on the posterior distribution, instead of {\tt zpeak}, to overcome relatively few numbers of filters in these fields. Scatter is defined by $\delta z = (z_{\rm phot}-z_{\rm spec})/(1+z_{\rm spec})$. Objects with more than 3 filters (red circles; $N=98$) are used for calculating the median of scatter, $\sigma_{\delta_z} \sim 0.12$ (dashed lines), while the rest are also shown ($N=94$; gray dots).
	}
\label{fig:specz}
\end{figure}

\begin{figure*}
\centering
	\includegraphics[width=1\textwidth]{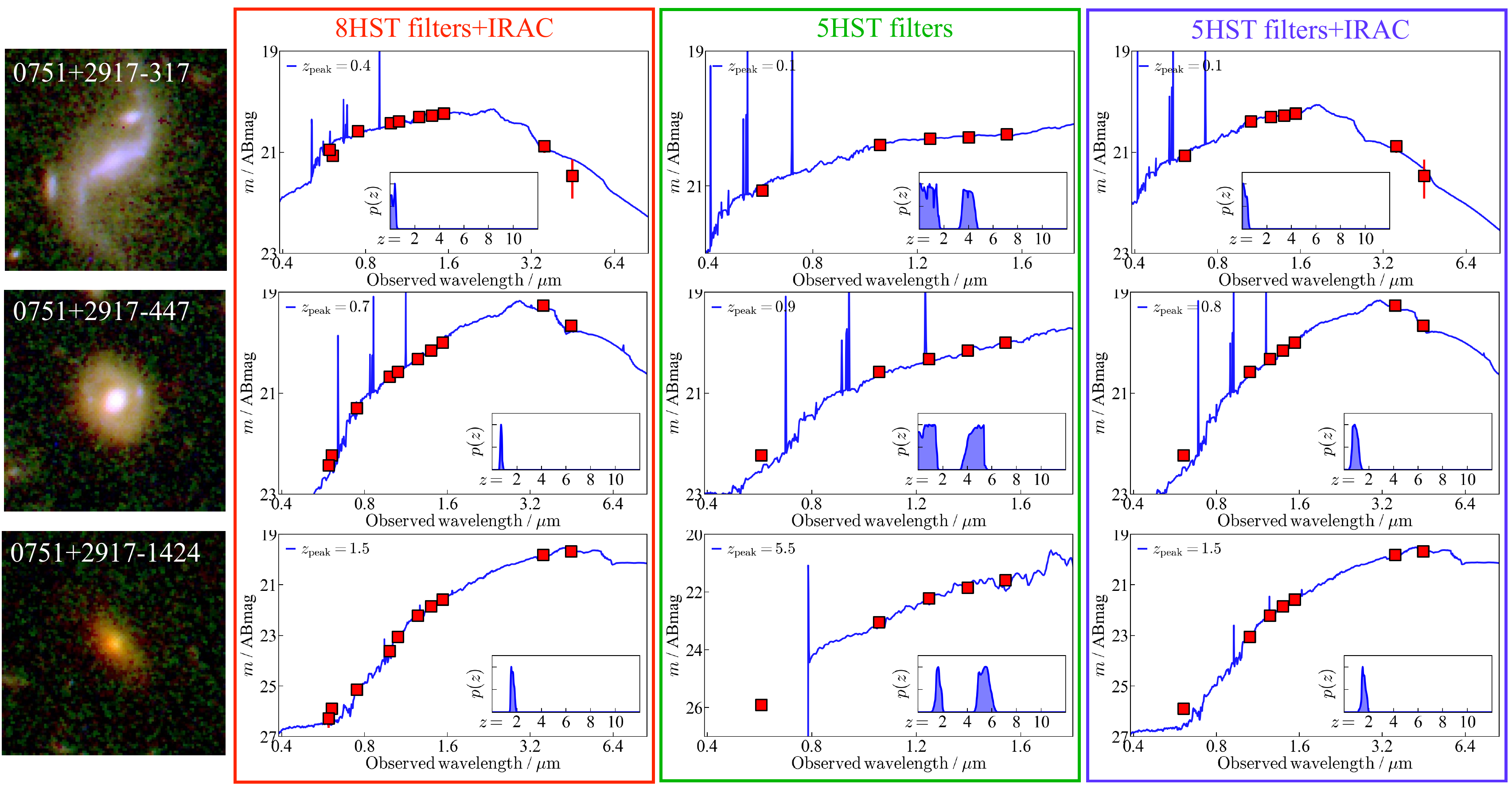}
	\caption{Example of three low-$z$ galaxies from the sBoRG-0751+2917 field, where a sufficient number of filters are available for a phot-$z$ comparison test. Each row shows results of individual galaxies, with a full set of filters (left), 5\,\hst\ filters as a fiducial example for BoRG (middle), and 5\,\hst\ filters and 2\,\spit\ filters (right). Redshift probability distributions are shown in the inset. Multiple redshift peaks seen in the results with the \hst-only filter set are resolved by adding \spit\ filters. 
	}
\label{fig:sed}
\end{figure*}

\section{Analyses}\label{sec:analysis}
\subsection{Photometric Redshift}\label{ssec:photz}
We derive photometric redshifts by using \eazy\ \citep{brammer08}, with its default template set (v1.3) and configuration parameters. The fitting redshift range is set to $z\in[0.01,12]$ with a step size of $\Delta \log(z+1)=0.01$. We use the default prior provided by \eazy, where brighter sources in F160W are more likely to be at lower redshift \citep[see also][]{morishita17}. Those who wish different configuration setups can use the flux catalog we provide and repeat the analysis.

We calculate absolute UV magnitude ($M_{\rm UV}$) and rest-frame colors by using the best fit SED at $z_{\rm peak}$. $M_{\rm UV}$ is calculated at rest-frame 1450\,\AA. Rest-frame colors, $U-V, B-V, V-J, z-J$, are calculated by convolving the best fit SED with filter response curves of standard filters.

While most of the \spbg\ fields are not covered spectroscopically, an exception is the CLASH program. Several fields of CLASH have spectroscopic follow-up observations, through the VLT-CLASH campaign (Sec.~\ref{sssec:data_clash}). In Fig.~\ref{fig:specz}, we compare our photometric redshifts to spectroscopic ones taken from the catalogs published by the CLASH team. Since most of the spectroscopic sources are relatively bright galaxies at $z<1$, we here use {\tt zmc} for photometric redshift, which is calculated from the posterior probability distribution with the F160W magnitude prior. The scatter is measured by $\delta z = (z_{\rm phot}-z_{\rm spec})/(1+z_{\rm spec})$, where we have $\langle \delta z \rangle=0.12$ for \Nspec\ objects with reliable spectroscopic redshift ($>80\%$) {\it and} a sufficient number of filters ($>3$) for photometric redshift calculation. Adopting $z_{\rm peak}$ instead of {\tt zmc} returns a larger scatter and higher fraction of catastrophic outliers, as the data set cannot distinguish, e.g., $z\sim1$ and $z\sim4$ solutions without the prior information in many cases (see below).

Since most of the CLASH fields typically have fewer filters than other programs, we also show an example of photometric redshift analysis on galaxies taken from the sBoRG-0751+2917 field, where 8\,\hst+IRAC filters are available (Fig.~\ref{fig:sed}). The field was originally observed in the cycle 17 HIPPIES program, and then followed up by \spit\ with IRAC ch1 and 2. We select three low-$z$ galaxies with different features; blue spiral galaxy, green disk+bulge galaxy, and passively evolving galaxy. With the full filter set, photometric redshift is well constrained, with a single peak in its probability distribution in all three cases. It is also noticed that \spit\ photometry plays an important role by constraining rest-frame $1.5\mu{\rm m}$ i.e. dust reddening and emission from low-mass stars; as a result, the derived redshift distribution has a single peak in all cases. 

To test photometric redshift quality of the same objects with reduced numbers of filters, we simply repeat the redshift fitting process for the same galaxies with the following two setups; 1.~one with only 5\,\hst\ filters (F350LP, F105W, F125W, F140W, and F160W), a default filter set for many of the \spbg\ fields, and 2.~the same 5\,\hst\ filters plus 2\,IRAC filters. The results are shown in Fig.~\ref{fig:sed}.

With only \hst\ filters, the probability distribution shows multiple peaks, which can be attributed to the fact that the set of filters only covers rest-frame $\sim0.4$ to $1\mu{\rm m}$ wavelength range and cannot distinguish degeneracy between the Balmer break at $z\sim1$ and the Lyman break $z\sim4$. Fields with a few number of \hst\ filters therefore may need additional setups e.g., by including redshift priors based on source morphology \citep{coe06}, luminosity \citep{brammer08} and/or environment \citep{morishita17}, when used for low-$z$ science cases. Adding IRAC photometry, on the other hand, returns a singly peaked redshift distribution, by constraining rest-frame $\sim2\mu{\rm m}$ wavelength range. The distribution is already similar to, but slightly broader than, the one derived with the full filter set.

\begin{figure}
\centering
	\includegraphics[width=0.47\textwidth]{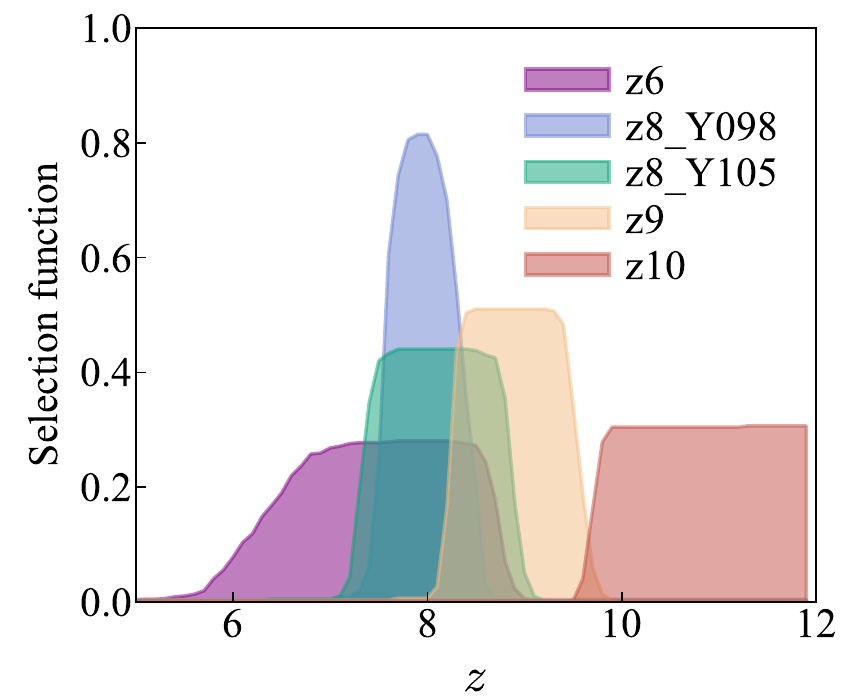}
	\caption{
	Effective redshift windows of the photometric color selections presented in Sec.~\ref{ssec:cs}. Distributions are normalized to an arbitrary constant.
	}
\label{fig:sele}
\end{figure}

\subsection{Color selections for high-z candidates}\label{ssec:cs}
To select high-$z$ source candidates, we follow color selection criteria defined in previous studies of the BoRG collaboration \citep{schmidt14,calvi16}. The color selection method is based on the Lyman break technique \citep{steidel96}, with some additional color cuts to effectively exclude low-$z$ interlopers. Signal-to-noise ratios and colors used in the following criteria are calculated with the aperture flux of $0.\!\arcsec64$ diameter.

In addition to color-cut criteria presented in the following subsections, we apply the following photometric cut uniformly to all high-$z$ source candidates;
$$S/N_{\rm non-detection\ filters}<1.0$$
$$S/N_{\rm min., non-detection\ filters}<1.0$$
$${\rm f_{persistent} = 0}$$
Non-detection filters are defined in each redshift range below, and objects without any non-detection filter coverage are excluded from the candidate list. In addition to the non-detection measured in a $0.\!\arcsec64$ diameter aperture ($S/N_{\rm non-detection\ filters}$), we measure $S/N$ in the non-detection filters with a smaller aperture ($0.\!\arcsec16$ diameter) to secure non-detection ($S/N_{\rm min., non-detection\ filters}$). For the same reason described in Sec.\ref{sssec:aper}, having unnecessarily large apertures can underestimate S/N and falsely identify an object as non-detection. 

${\rm f_{persistent}}$ is a flag for IR persistence. Specifically, for each of selected candidates, we visit IR persistent model images created for each visit of the field, and check if the source position is affected by persistence.\footnote{\url{https://archive.stsci.edu/prepds/persist/search.php}} If there is any overlap between the segmentation map of the source and the flagged area in the IR persistence model, the object is flagged (${\rm f_{persistent}}=1$) and excluded from the candidate list.

\subsubsection{$z\sim10$ galaxy candidates}
$$S/N_{\rm 160}>6.0$$
$$J_{125}-H_{160}>1.3$$
$$H_{160}-[3.6]<1.4\ {\rm (when\ IRAC\ ch1\ is\ available)}$$
(z10 selection). Non-detection filters are F105W and bluer filters. The additional $H_{160}-[3.6]<1.4$ cut, only when IRAC ch1 is available, offers more robust candidates \citep{bouwens15}.

\subsubsection{$z\sim9$ galaxy candidates}
$$S/N_{\rm 140}>6.0 \land$$
$$S/N_{\rm 160}>4.0\land$$
$$Y_{105}-JH_{140}>1.5\land$$
$$JH_{140}-H_{160}<0.3\land$$
$$Y_{105}-JH_{140}>5.33\cdot(JH_{140}-H_{160})+0.7$$
(z9 selection). Non-detection filters are F098M and bluer filters.

\subsubsection{$z\sim8$ galaxy candidates}\label{sec:sample}
For fields where F105W filter is available,
$$S/N_{\rm 125}>6.0\land$$
$$S/N_{\rm 160}>4.0\land$$
$$Y_{105}-J_{125}>0.45\land$$
$$J_{125}-H_{160}<0.5\land$$
$$Y_{105}-J_{125}>1.5\cdot(J_{125}-H_{160})+0.45$$
(z8\_Y105 selection). Non-detection filters are F814W and bluer filters.

For fields where F098M filter is available,
$$S/N_{\rm 125}>6.0\land$$
$$S/N_{\rm 160}>4.0\land$$
$$Y_{098}-J_{125}>1.75\land$$
$$J_{125}-H_{160}<0.5\land$$
$$(J_{125}-H_{160})<0.02+0.15(Y_{098}-J_{125}-1.75)$$
 (z8\_Y098 selection). Non-detection filters are F814W and bluer filters.

\subsubsection{$z\sim6$ galaxy candidates}\label{sec:sample}

For fields where F814W filter is available,
$$S/N_{\rm 125}>6.0\land$$
$$S/N_{\rm 160}>4.0\land$$
$$I_{814}-J_{125}>0.8\land$$
$$J_{125}-H_{160}<0.4\land$$
$$I_{814}-J_{125}>2.0\cdot(J_{125}-H_{160})+0.8\land$$
(z6 selection). Non-detection filters are F606W and bluer filters. It is noted that, as shown in Fig.~\ref{fig:sele}, this selection has a wide selection window in the redshift space, partially overlapping with other selections at $z\sim7$ to 8. However, having this selection is still effective in a field where no $Y$-band is available, which is occasionally the case for some of the BoRG cycle 25 and CLASH fields. We still keep the same labeling, ``z6", as in the original reference of this selection \citep{bouwens15}, whereas the median of the probability distribution is $z\sim7.5$.

\begin{figure*}
\centering
	\includegraphics[width=0.8\textwidth]{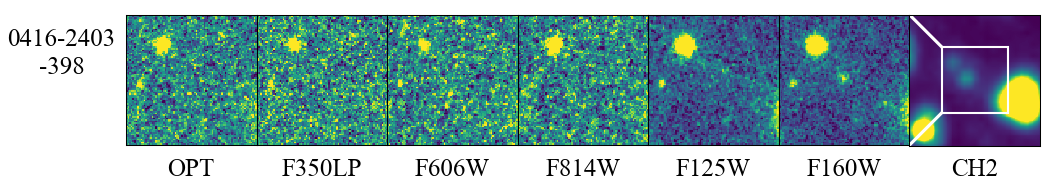}
	\includegraphics[width=0.5\textwidth]{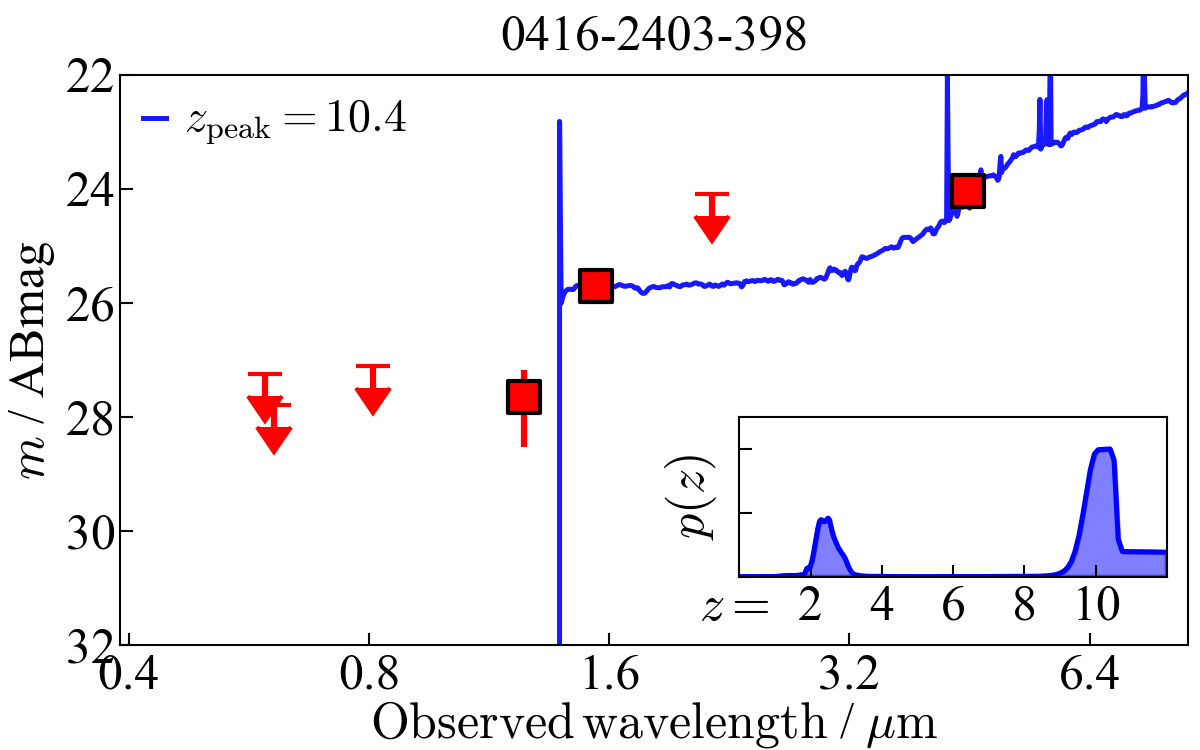}
	\caption{High-redshift source candidate sBoRG-0416-2403-398 identified in the z10 selection. The redshift probability distribution peaks at $z\sim10.4$ with $M_{UV}=-21.9$\,mag, characterizing this as one of the most luminous candidates at $z\sim10$ in the literature. Significant detection in IRAC ch2, however, implies that the object could be a $z\sim3$ dusty galaxy instead, with $p(z<6)=23\%$. Postage stamps are $6.\!\arcsec5 \times 6.\!\arcsec5$ in size for \hst\ and $12.\!\arcsec8 \times 12.\!\arcsec8$ for \spit\ ch2. The size of the \hst\ stamps is shown in the \spit\ image (white rectangle). OPT is for a stacked image of optical bands (F814W and bluer filters).
	}
\label{fig:highz}
\end{figure*}

\begin{figure*}
\centering
	\includegraphics[width=0.9\textwidth]{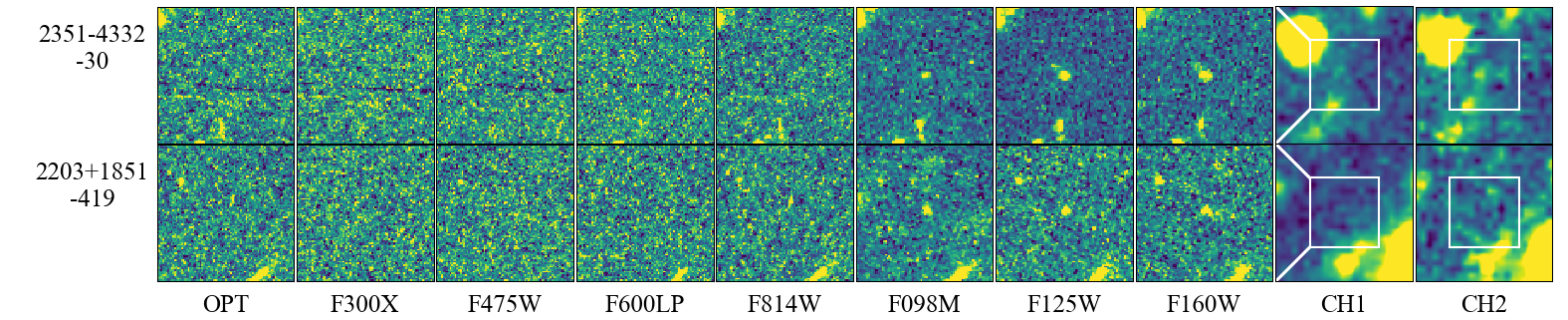}
	\includegraphics[width=0.45\textwidth]{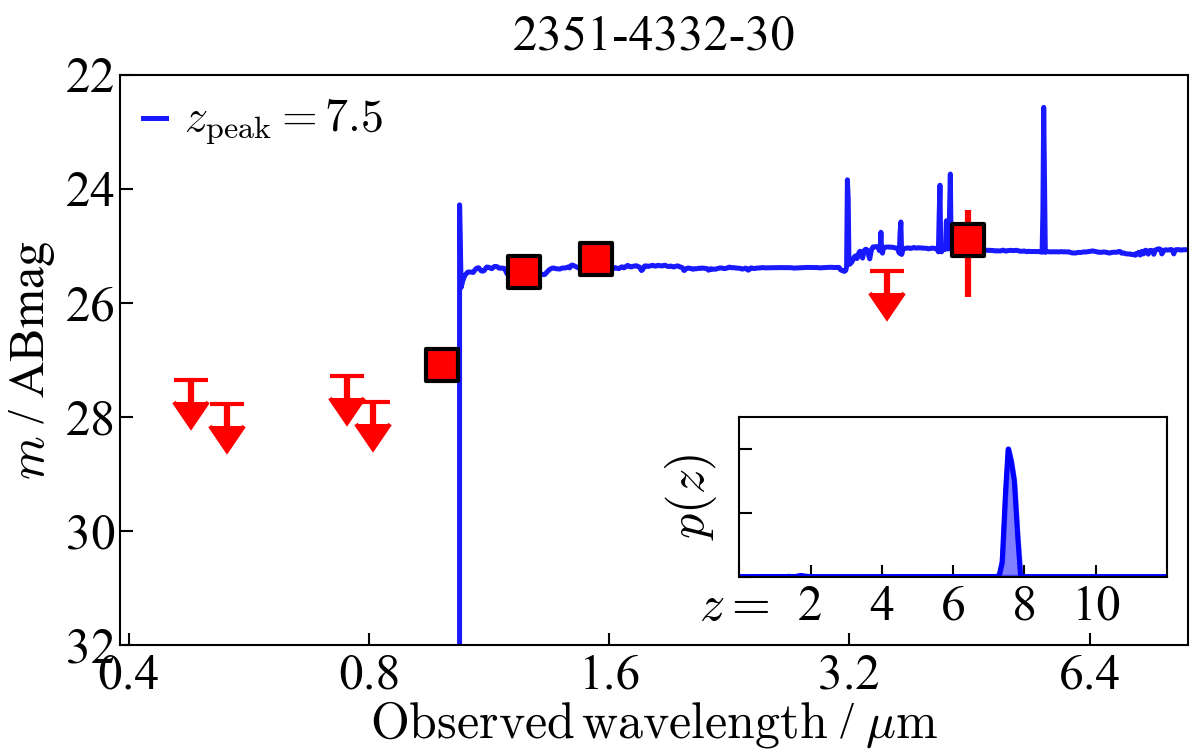}
	\includegraphics[width=0.45\textwidth]{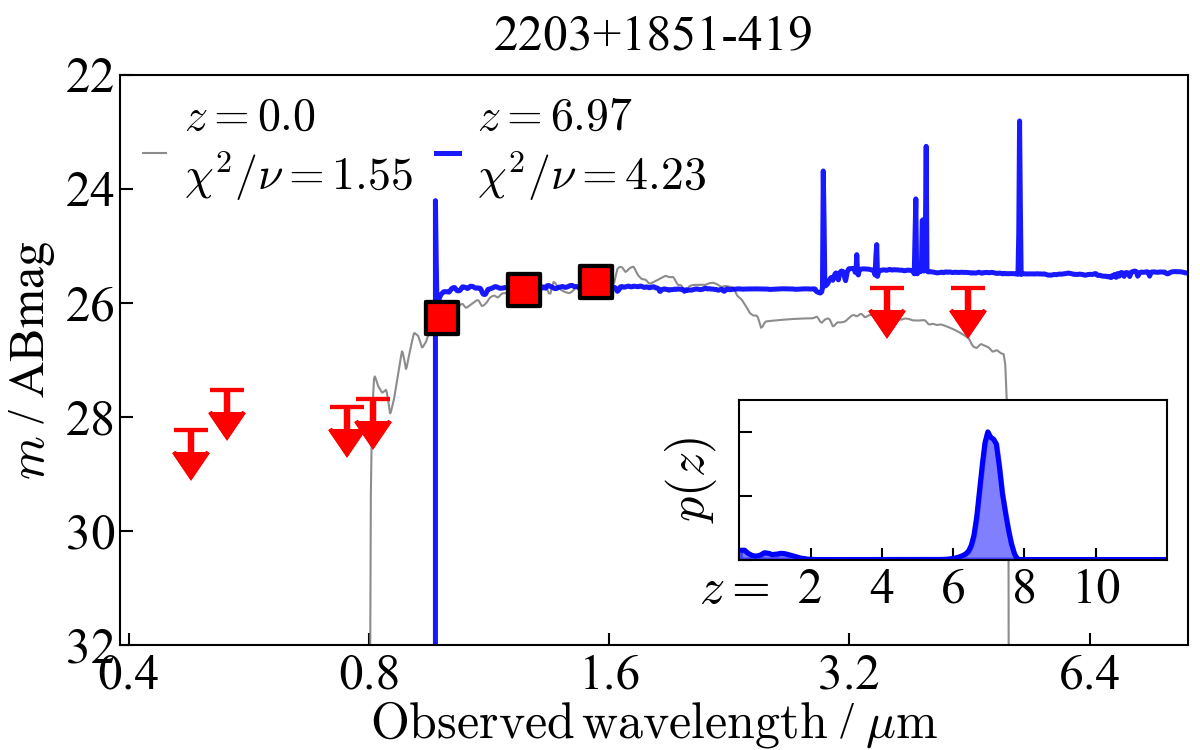}
	\caption{
	Same as Fig.~\ref{fig:highz} but for two source candidates selected in the z6 selection. Clean non-detection in the optical filters and tentative/non-detection in the two IRAC bands make the photometric redshift strongly peaked at $z\simgt7$. sBoRG-2203+1851-419 has a point-source morphology; SED fitting analysis with a set of dwarf templates indicates that the object is more likely a foreground brown dwarf (gray solid line). The analysis with the dwarf templates is not reflected in the probability distribution inset.
	}
\label{fig:highz2}
\end{figure*}

\subsubsection{Caveats on High-$z$ Source Color Selection}\label{sssec:cav}
Redshift distributions for the color selections above are shown in Fig.~\ref{fig:sele}. These selection functions are estimated in a completeness simulation in the same manner as in \citet[][]{morishita18b}. It is noted that the shape of each selection function may slightly be changed depending on the depth of filters used for the selection, while the purpose here is to provide an approximate redshift range for each color selection. A dedicated study will provide completeness and contamination rates of high-$z$ source candidates (Leethochawalit, in prep.).

There are two caveats that interested readers may consider: besides the color cuts, combining photometric redshift results can improve the selection and exclude low-$z$ interlopers. For example, \citet{morishita18b} only selected those with probability distribution at $z<6.5$ smaller than $30\%$, effectively excluding suspicious candidates. However, it is also noted that having too stringent conditions could exclude genuine high-$z$ galaxies with e.g., a moderate amount of dust, as their SEDs are impossible to distinguish from those of low-$z$ interlopers with the filter sets used here, leading to an {\it incomplete} high-$z$ candidate list. The other note is that selected candidates may still need visual inspection, to exclude contaminations such as stellar spikes, false detection of noise at the edge of images, false non-detection images i.e. those falling in the detector gap of WCF3-UVIS and ACS (whereas such a gap does not exist in the WFC3-IR channel). Lastly, the color selection criteria presented below are not exclusive of each other, and may be classified in more than one selection, as shown in Fig.\ref{fig:sele}. Such duplication of redshift range is inevitable for more complete candidate selection with a dataset like \spbg, where multiple filter configurations are available.

\subsection{Example of Selected High-$z$ Candidates}\label{sssec:res}
As a demonstration of high-$z$ source selection and analysis, we here present newly identified source candidates from two programs, COS-GTO and CLASH, from 22 and 44 fields, respectively. These fields only allow the z10, z8\_Y098, and z6 selections, while identified sources by the z8\_Y098 selection were initially presented in \citet{bradley12} and \citet{schmidt14}. We therefore focus on the z10 selection and z6 selection in this study, as the purpose is to demonstrate how the actual selection proceeds as well as to provide a few caveats. For the rest of the high-$z$ source candidates selected from the entire \spbg\ fields, and for comparison with the previously identified sources, readers are referred to a forthcoming paper (Roberts-Borsani, in prep.). 

After applying the z10 selection above and $p_{\rm high}>0.7$ (see Table~\ref{tab:col}), we found 59\,sources. Most of the sources (54) are from the CLASH program. However, our visually inspection found a significantly high contamination fraction, where only three of them are found as possible high-$z$ sources; the rest are either stellar spikes, noise structures at the edge of the detectors, or sources with suspicious non-detection where positive pixels are observed near the position in the optical images. This high contamination fraction in the CLASH fields is attributed to a relatively small number of available filters ($3$), while this is not the case for the COS-GTO data or other previous BoRG fields \citep[e.g.,][]{calvi16,morishita18b}.

Two of the three possible candidates have significant IRAC ch1 detection, with F160W$-[3.6]>1.4$, which are rejected by an additional criterion for the z10 selection. The other one, sBoRG-0416-2403-398, does not have ch1 coverage but ch2, and is detected at $S/N=6.4$. While the color is still red (F160W$-[4.5]=1.8$), this does not immediately exclude the object from the z10 list (Fig.~\ref{fig:highz}).

We check availability of other IR filters. A moderately deep $K_S$-band image taken with VISTA/VIRCAM is available (PID:198.A-2008, PI. Nonino, M.). We do not find any detection at the position of sBoRG-0416-2403-398, and thus we place an upper limit listed in the ESO-archive ($K_S=22.34$\,ABmag, $5\sigma$). Placing this upper limit does not change the result, and its redshift probability distribution still peaks at $z\sim10$. 

Given the cosmic time at the inferred redshift, it is unlikely that this object has a very mature population of this color, implying a moderate amount of dust in the system is needed; such examples are seen in recent observations by ALMA \citep{watson15,laporte17,tamura18}. We fit the observed data points with a bayesian SED fitting code, {\tt gsf} \citep{morishita19} with a setup of fsps \citep{conroy09fsps} stellar population templates of solar metallicity, \citet{chabrier03} IMF, and \citet{calzetti00} dust attenuation while fixing redshift to $z=10.4$, and the best-fit result infers $A_V=1.4_{-0.3}^{+0.2}$, with $\logm= 10.0_{-0.3}^{+0.8}$.

On the other hand, because of the significant ch2 flux, the source also has a secondary redshift peak at $z\sim3$. The total probability of a lower redshift ($z<6$) solution is $\sim23\%$. To better constrain its SED and photometric redshift, additional photometric data points at $\sim2\,\mu {\rm m}$ to $\sim4\,\mu {\rm m}$ are desired, while spectroscopic follow-up of rest-frame UV emission lines would be challenging from the current ground facilities given its apparent magnitude.

Three fields in the COS-GTO program allow the z6 selection. Three sources are identified by the z6 selection criteria, one of which is rejected by visual inspection for possible flux contaminations in the non-detection images.
The other two candidates have a sufficient number of filters (Fig.~\ref{fig:highz2}), including two IRAC bands, that constrains their photometric redshift well. sBoRG-2351-4332-30 has a striking color break ($\mathrm{F098M-F125W}=1.6$), making its redshift probability peak at $z\sim7.5$. We see clean non-detection in the 4 optical filters. Its IRAC ch2 image detects moderate flux ($24.9\pm0.6$\,mag) without being significantly contaminated by neighboring objects while there is no detection in ch1. The object has a flat UV slope, effectively eliminating any significant possibility as a low-$z$ interloper ($<1\%$).

The other z6 candidate, sBoRG-2203+1851-419, also has clean non-detection in the optical bands. Fluxes are not detected in moderately deep IRAC images, placing upper limits to its SED. In combination with a flat $J_{125}-H_{160}$ color and the IRAC non-detection, the object has a small probability as a low-$z$ interloper ($\sim12\%$).

{Furthermore, the object is characterized by a compact morphology, and indeed satisfies the criteria for point-source selection defined by Eq.~\ref{eq:ps} in Section~\ref{ssec:size}.
Such high-$z$ sources with point-source morphologies are of particular interest, with regard to their possibility as low-luminosity quasars or compact starburst galaxies. \citet{morishita20} identified three candidates by the z8\_Y105 selection and investigated their properties.}

We repeat the photometric analysis for sBoRG-2203+1851-419 but this time with a set of dwarf templates taken from the IRTF spectral Library \citep{rayner03}, in the same way as in \citet{morishita20}. The best-fit result provides a considerably smaller reduced-$\chi^2$ value (1.6 cf. 4.23 with extragalactic templates; see Fig.~\ref{fig:highz}). This is not a surprise given a considerably high fraction of contaminating cool stars at this redshift, in contrast to the one at $z\simgt8$ \citep[e.g., Fig.~4 of][]{morishita20}. We therefore conclude this object is a foreground star. The sources presented here are summarized in Table~\ref{tab:photom}.

\section{Output products}\label{sec:products}
As output products, we provide three separate catalogs --- a master catalog that contains basic source properties such as coordinates, extraction flag, and structural parameters measured by \sext\ (Table~\ref{tab:master}), a flux catalog that lists flux measurements (Table~\ref{tab:phot}), and a redshift catalog that contains spectroscopic and photometric redshifts, absolute UV magnitude ($M_{\rm UV}$) and rest-frame colors calculated with the best-fit SED (Table~\ref{tab:eazy}). All catalogs contain common objects that are brighter than the $3\sigma$ limiting magnitude in F160W of each field ($N=\Nobj$). The columns included in the catalogs are summarized in Table~\ref{tab:col} with brief descriptions. The extraction flag in the master catalog is derived by \sext, a sum of powers of 2 coded in decimal. For example, the flag can be used to check if a source has flux contamination from neighboring sources ($+1$), is de-blended from neighboring sources ($+2$), or is saturated ($+4$).\footnote{\url{https://sextractor.readthedocs.io/en/latest/Flagging.html}} Photometric flux measurements listed in the flux catalog are corrected for the RMS scaling (Sec.~\ref{ssec:rms}) and dust attenuation (Sec.~\ref{ssec:photom}), and scaled so they have a common magnitude zeropoint $m_0$ of 25 i.e. $m=-2.5 \log f+m_0$\,ABmag. 

{\hst\ imaging data reduced in this study will be released with the final pixel scale of $0.\!''08$, which include co-added science frames, scaled RMS maps, and segmentation map, for all fields.

These catalogs, reduced imaging data (DOI: 10.17909/t9-m7tx-qb86), and \sext\ parameter files used for the flux measurement will be available via the dedicated site\footnote{\url{https://www.stsci.edu/~tmorishita/superborg/}} and the Mikulski Archive for Space Telescopes (MAST).
}

\section{Summary}\label{sec:sum}
In this study, we introduced a new archival project, \spbg, which collects previous \hst\ parallel imaging data taken for extragalactic science cases, including pure-parallel (BoRG, HIPPIES, and COS-GTO) and coordinated-parallel programs (CLASH and RELICS). The final data set consists of the total number of \Nfld\ independent sightlines, reaching the total effective area of $\sim$\,\Afld, which is the largest-area extragalactic survey with \hst\ for extragalactic science. We reduced these datasets in a consistent way, with an updated version of our customized pipeline. We showed that our new sky background subtraction step improves the limiting magnitude by $\sim0.1$\,mag, detecting a few hundreds more faint sources at $\sim27$\,mag per field. We also showed that our revise in the photometric flux extraction method, from previous isophotal to aperture, can minimize possible false identification of low-$z$ sources as high-$z$ source candidates. 

We presented photometric analyses, including photometric redshift analysis and color selection of high-$z$ source candidates by means of the Lyman break technique. In the last section, we presented newly identified luminous source candidates at $z\simgt7$ from the fields of two programs as a preliminary result. One of them has a primary redshift peak at $z\sim10.4$, though its strong detection in IRAC ch2 suggests that this may be a low-$z$ dusty galaxy ($\sim23\%$). The other two candidates have a strong redshift peak $z\sim7$, while one of them has a point-source morphology; phot-$z$ analysis with a set of dwarf templates indeed inferred that this object is more likely a foreground star. A complete list and further analysis of high-$z$ source candidates from the entire \spbg\ fields will be presented in a forthcoming paper.

While the primary focus of \spbg\ is on luminous source candidates at high redshift, the dataset is also suitable for science cases at low and intermediate redshifts, especially in fields with supplemental data from follow-up \hst\ and \spit\ observations. The imaging data products and catalogs will be publicly available.

\acknowledgments
The author thanks the anonymous referee for carefully reading the manuscript and providing constructive comments. 
The author is grateful to the BoRG, HIPPIES, COS-GTO, CLASH, and RELICS collaborations for their significant effort in designing their parallel opportunities valuable and publicly available. The author is grateful to Michele Trenti for initiating the BoRG program and providing strategic details of its observations, Larry Bradley and Austin Hoag for constructing the basis of the BoRG reduction pipeline, Massimo Stiavelli and Ray Lucas for carefully reading the manuscript and providing useful comments, Tommaso Treu, Guido Roberts-Borsani, and Nicha Leethochawalit for kind support and fruitful discussion, and Marc Postman for providing details on the parallel data of the CLASH program. Support for this work was provided by NASA through grant numbers HST-GO-15212.002, HST-GO-15702.002, and HST-AR-15804.002-A from the Space Telescope Science Institute, which is operated by AURA, Inc., under NASA contract NAS 5-26555.

\noindent
{\it Software:} Astropy \citep{astropy:2013, astropy:2018}, numpy \citep{oliphant2006guide,van2011numpy}, python-fsps \citep{foreman14}, T-PHOT \citep{merlin16}, gsf \citep{morishita19}, fsps \citep{conroy09fsps}, EAzY \citep{brammer08}, LACOSMIC \citep{vandokkum01}, Astrodrizzle \citep{hack12}, SExtractor \citep{bertin96}, Pypher \citep{boucaud16}, GALFIT \citep{peng02, peng10}.

\clearpage


\tabletypesize{\small}
\tabcolsep=2pt
\startlongtable
\begin{deluxetable*}{ccccccccccccccc}
\renewcommand{\arraystretch}{0.9}
    \tablecolumns{15}
    \tablewidth{0pt}
    \tablecaption{}
\tablehead{
    \colhead{Field} &
    \colhead{Survey} &
    \colhead{R.A.} &
    \colhead{Decl.} &
    \colhead{F350LP} &
    \colhead{F600LP} &
    \colhead{F606W} &
    \colhead{F814W} &
    \colhead{F098M} &
    \colhead{F105W} &
    \colhead{F125W} &
    \colhead{F140W} &
    \colhead{F160W} &
    \colhead{IRAC1} &
    \colhead{IRAC2}
\vspace{-0.3cm}\\
    \colhead{} &
    \colhead{} &
    \colhead{deg} &
    \colhead{deg} &
    \colhead{mag} &
    \colhead{mag} &
    \colhead{mag} &
    \colhead{mag} &
    \colhead{mag} &
    \colhead{mag} &
    \colhead{mag} &
    \colhead{mag} &
    \colhead{mag} &
    \colhead{mag} &
    \colhead{mag}
}
\startdata
\cutinhead{BoRG Cycle17}
0540-6409 & B17 & 8.488e+01 & -6.415e+01 & --- & --- & 26.88 & --- & 26.54 & --- & 26.49 & --- & 26.17 & --- & --- \\
0553-6405 & B17 & 8.827e+01 & -6.409e+01 & --- & --- & 27.04 & --- & 26.78 & --- & 26.66 & --- & 26.24 & 23.29 & 23.38 \\
0624-6432 & B17 & 9.590e+01 & -6.453e+01 & --- & --- & 27.03 & --- & 26.26 & --- & 26.23 & --- & 25.64 & --- & --- \\
0624-6440 & B17 & 9.595e+01 & -6.466e+01 & --- & --- & 26.86 & --- & 26.57 & --- & 26.37 & --- & 25.90 & --- & 23.34 \\
0637-7518 & B17 & 9.927e+01 & -7.531e+01 & --- & --- & 27.12 & --- & 26.74 & --- & 26.51 & --- & 25.88 & 23.62 & 23.27 \\
0756+3043 & B17 & 1.190e+02 & 3.072e+01 & --- & --- & 26.98 & --- & 26.76 & --- & 26.57 & --- & 26.11 & --- & --- \\
0808+3946 & B17 & 1.221e+02 & 3.976e+01 & --- & --- & 26.82 & --- & 26.57 & --- & 26.47 & --- & 25.94 & 23.45 & 23.00 \\
0819+4911 & B17 & 1.248e+02 & 4.918e+01 & --- & --- & 26.71 & --- & 26.66 & --- & 26.36 & --- & 25.73 & --- & --- \\
0820+2332 & B17 & 1.250e+02 & 2.354e+01 & --- & --- & 26.89 & --- & 26.51 & --- & 26.09 & --- & 25.84 & --- & --- \\
0906+0255 & B17 & 1.364e+02 & 2.926e+00 & --- & --- & 27.08 & --- & 26.94 & --- & 26.90 & --- & 26.48 & 23.95 & 23.24 \\
0909+0002 & B17 & 1.373e+02 & -3.023e-02 & --- & --- & 28.25 & --- & 27.95 & --- & 27.79 & --- & 26.94 & 23.69 & 23.40 \\
0922+4505 & B17 & 1.404e+02 & 4.509e+01 & --- & --- & 26.81 & --- & 26.68 & --- & 26.56 & --- & 26.21 & --- & --- \\
1031+3804 & B17 & 1.577e+02 & 3.806e+01 & --- & --- & 26.65 & --- & 26.37 & --- & 26.27 & 25.30 & 25.96 & 24.17 & 23.31 \\
1152+5441 & B17 & 1.780e+02 & 5.468e+01 & --- & --- & 27.24 & --- & 27.01 & --- & 26.96 & 26.19 & 26.60 & 23.67 & --- \\
1153+0056 & B17 & 1.782e+02 & 9.319e-01 & --- & --- & 27.02 & --- & 26.76 & --- & 26.72 & --- & 26.33 & 24.17 & 23.35 \\
1230+0750 & B17 & 1.875e+02 & 7.826e+00 & --- & --- & 26.72 & --- & 26.08 & --- & 25.83 & --- & 25.60 & 23.40 & --- \\
1245+3356 & B17 & 1.912e+02 & 3.394e+01 & --- & --- & 26.88 & --- & 26.78 & --- & 26.63 & --- & 26.06 & 24.07 & 23.53 \\
1437+5043 & B17 & 2.192e+02 & 5.072e+01 & --- & --- & 26.92 & --- & 27.31 & 27.05 & 27.26 & --- & 26.97 & 23.93 & 23.40 \\
1632+3733 & B17 & 2.481e+02 & 3.756e+01 & --- & --- & 27.01 & --- & 26.89 & --- & 26.83 & 25.67 & 26.45 & 23.73 & 23.32 \\
1632+3737 & B17 & 2.479e+02 & 3.761e+01 & 28.00 & --- & 26.74 & --- & 26.80 & 27.38 & 27.33 & 27.42 & 27.12 & 23.67 & 23.34 \\
\cutinhead{BoRG Cycle19}
0456-2203 & B19 & 7.396e+01 & -2.205e+01 & --- & --- & 27.02 & --- & 26.71 & --- & 26.64 & 24.89 & 26.36 & 23.78 & 23.30 \\
0951+3304 & B19 & 1.477e+02 & 3.307e+01 & --- & --- & 26.66 & --- & 26.47 & --- & 26.40 & --- & 26.15 & 24.00 & 23.21 \\
0952+5304 & B19 & 1.479e+02 & 5.307e+01 & --- & --- & 27.02 & --- & 26.68 & --- & 26.72 & --- & 26.44 & --- & --- \\
1059+0519 & B19 & 1.647e+02 & 5.312e+00 & --- & --- & 26.79 & --- & 26.70 & --- & 26.65 & --- & 26.40 & 24.02 & 23.28 \\
...&...&...&...&...&...&...&...&...&...&...&...&...&...&...\\
0722+0729 & REL & 1.105e+02 & 7.484e+00 & 26.30 & --- & --- & --- & --- & 26.22 & 26.12 & 26.11 & 25.98 & --- & --- \\
0830+6555 & REL & 1.276e+02 & 6.591e+01 & 26.75 & --- & --- & --- & --- & 26.44 & 26.47 & 26.67 & 26.26 & 23.50 & 23.21 \\
0843+3627 & REL & 1.307e+02 & 3.646e+01 & 26.02 & --- & --- & --- & --- & 26.21 & 26.26 & 26.20 & 26.10 & 23.86 & 23.38 \\
1132-1950 & REL & 1.729e+02 & -1.983e+01 & 26.73 & --- & --- & --- & --- & 26.52 & 26.30 & 26.34 & 26.24 & 23.81 & 23.63 \\
1335+4054 & REL & 2.038e+02 & 4.090e+01 & 26.99 & --- & --- & --- & --- & 26.75 & 26.43 & 26.52 & 26.40 & 24.18 & 24.36 \\
1515-1517 & REL & 2.287e+02 & -1.528e+01 & 26.61 & --- & --- & --- & --- & 26.42 & 26.24 & 26.30 & 26.15 & 23.32 & 23.04 \\
1521-8134 & REL & 2.302e+02 & -8.157e+01 & 26.77 & --- & --- & --- & --- & 26.39 & 26.52 & 26.59 & 26.30 & 23.23 & 23.27 \\
1917-3335 & REL & 2.894e+02 & -3.359e+01 & 26.66 & --- & --- & --- & --- & 26.42 & 26.27 & 26.31 & 26.13 & --- & --- \\
2212-0354 & REL & 3.330e+02 & -3.903e+00 & 26.81 & --- & --- & --- & --- & 26.45 & 26.30 & 26.29 & 26.19 & 23.82 & 23.82 \\
\enddata
\tablecomments{Example of the field table, listing coordinates and $5\,\sigma$-limiting magnitudes of filters. The full table is available online.}
\label{tab:field}
\end{deluxetable*}

\begin{deluxetable*}{lcccccccccc}
\renewcommand{\arraystretch}{0.9}
\tabletypesize{\small}
\tabcolsep=7pt
\tablecolumns{11}
\tablewidth{0pt}
\tablecaption{Photometric properties of high-$z$ source candidates identified in the CLASH and COS-GTO fields.}
\tablehead{
	\colhead{Field ID} &
	\colhead{Survey} &
	\colhead{ObjID} & 
	\colhead{R.A.} & 
	\colhead{Decl.} & 
	\colhead{$z_{\rm peak}$}  & 
	\colhead{$M_{\rm UV}$}  & 
	\colhead{$r$} & 
	\colhead{$m_{160}$} & 
	\colhead{$\chi^2/\nu$} & 
	\colhead{${\chi^2/\nu}_{\rm dw}$}
\vspace{-0.3cm}\\
	\colhead{} & 
	\colhead{} & 
	\colhead{} & 
	\colhead{deg} & 
	\colhead{deg} & 
	\colhead{} & 
	\colhead{mag}  & 
	\colhead{kpc} & 
	\colhead{mag} & 
	\colhead{} & 
	\colhead{}
}
\startdata
\cutinhead{z10 selection}
sBoRG-0416-2403 & clash & 398 & 63.923630 & -24.034565 & 10.4 & -21.9 & 0.80 & 25.7 & 0.10 & 15.02\\
\cutinhead{z6 selection}
sBoRG-2351-4332 & cosgto & 30 & 357.644440 & -43.542427 & 7.55 & -21.7 & 0.98 & 25.2 & 3.60 & 9.32\\
sBoRG-2203+1851$^\dagger$ & cosgto & 419 & 330.706510 & 18.847076 & 6.97 & -21.3 & 1.03 & 25.6 & 4.23 & 1.55
\enddata
\tablenotetext{\rm {\bf Notes.}
}
{
\\
$z_{\rm peak}$ : Peak photometric redshift estimated with \eazy.
$\chi^2/\nu$ : Reduced chi-square from \eazy\ photometric redshift fitting analysis with extragalactic templates.
$\chi_{\rm dw}^2/\nu$ : Reduced chi-square with brown dwarf templates.\\
$\dagger$: This object has a point-source morphology, and in combination with the redshift fitting analysis we conclude that the object is most likely a foreground star (Sec.~\ref{sssec:res}).
}
\label{tab:photom}
\end{deluxetable*}

\begin{deluxetable*}{lcl}
\tabletypesize{\footnotesize}
\tabcolsep=6pt
\tablecolumns{3}
\tablewidth{0pt} 
\tablecaption{Columns and description for the photometric catalogs.}
\tablehead{
\colhead{Column name} & \colhead{Units} & \colhead{Description}
}
\startdata
\cutinhead{Master catalog}
field & & BoRG field name.\\
survey & & Survey name. B{\ttfamily xx}: BoRG cycle{\ttfamily xx}. H{\ttfamily xx}: HIPPIES cycle{\ttfamily xx}. CLA: CLASH. REL: RELICS. COS: COS-GTO.\\
id & & Individual identification name of objects.\\
ra &degree & R.A. (J2000).\\
dec &degree & Declination (J2000).\\
x &pixel & x position in image coordinate.\\
y &pixel & y position in image coordinate.\\
kron\_radius &pixel & Kron radius measured by \sext.\\
a\_image &pixel & Radius along the major axis measured by \sext.\\
b\_image &pixel & Radius along the minor axis measured by \sext.\\
theta &degree & Position angle measured by \sext.\\
class\_star & & Class star indicator measured by \sext.\\
flux\_radius &pixel & Non-parametric half-light radius measured by \sext.\\
flags & & Extraction flag from \sext.\\
f\_persist & & Flag for IR-image persistence at object positions. 0: Clean, 1: Contaminated.\\
flag\_detect & & Flag for source detection in F160W images. 0: $S/N<3$, 1: $S/N>3$.\\
flux\_ratio0 & & Flux concentration ratio $f_0/f_1$, where $f_0$ ($f_1$) is flux measured in an aperture of diameter size $2$ (4)\,pixels.\\
flux\_ratio1 & & Flux concentration ratio $f_1/f_2$, where $f_1$ ($f_2$) is flux measured in an aperture of diameter size $4$ (8)\,pixels.\\
flux\_scale & & Scale ratio of total flux to aperture flux measured in F160W. For those with scale $<1$, scale is set to 1.\\
faper\_F160W &$f_{\nu}$  & Aperture flux in F160W, with diameter size of 8\,pixels.\\
eaper\_F160W &$f_{\nu}$  & Flux error of faper\_F160W.\\
faper\_F140W &$f_{\nu}$  & Aperture flux in F140W, with diameter size of 8\,pixels.\\
eaper\_F140W &$f_{\nu}$  & Flux error of faper\_F140W.\\
\cutinhead{Flux catalog}
f\_{\ttfamily xx} &$f_{\nu}$  & Total flux of objects in band {\ttfamily xx}.\\
e\_{\ttfamily xx} &$f_{\nu}$  & Flux error ($1\sigma$) in total flux of objects in band {\ttfamily xx}.\\
flag\_{\ttfamily xx} & & Extraction flag for IRAC photometry by tphot. IRAC bands (channels 1 to 4) only.\\
fiso\_{\ttfamily xx} &$f_{\nu}$  & Isophotal flux of objects in band {\ttfamily xx}.\\
eiso\_{\ttfamily xx} &$f_{\nu}$ & Isophotal flux error ($1\sigma$) in total flux of objects in band {\ttfamily xx}.\\
\cutinhead{Redshift catalog}
zspec & & Spectroscopic redshift, if available.\\
flag\_zspec & & Spectroscopic redshift quality flag. $2$: LIKELY ($\sim80\%$ reliability), $3$: SECURE ($\sim100\%$), $9$: SINGLE-LINE ($>90\%$).\\
z16 & & Photometric redshift at the 16th percentile of redshift distribution.\\
z50 & & Photometric redshift at the 50th percentile of redshift distribution. \\
z84 & & Photometric redshift at the 84th percentile of redshift distribution. \\
zpeak & & Photometric redshift at the peak of redshift distribution.\\
MUV16 &mag & Absolute UV (1450\,\AA) magnitude calculated at z16.\\
MUV50 &mag & Absolute UV (1450\,\AA) magnitude calculated at z50.\\
MUV84 &mag & Absolute UV (1450\,\AA) magnitude calculated at z84.\\
MUVpeak &mag & Absolute UV (1450\,\AA) magnitude calculated at zpeak.\\
zset & & A limit redshift for the calculation of plow and phigh.\\
plow & & Total probability of redshift distribution at $z<{\rm zset}$.\\
phigh & & Total probability of redshift distribution at $z>{\rm zset}$.\\
UVbeta\_lambda & & UV-beta slope ($\beta_\lambda$) calculated with the best-fit template at zpeak.\\
U-V &mag & Rest-frame $U-V$ color calculated with the best-fit template at zpeak.\\
B-V &mag & Rest-frame $B-V$ color calculated with the best-fit template at zpeak.\\
V-J &mag & Rest-frame $V-J$ color calculated with the best-fit template at zpeak.\\
z-J &mag & Rest-frame $z-J$ color calculated with the best-fit template at zpeak.\\
chi2 & & Reduced chi-square at zpeak.\\
zmc & & Photometric redshift derived with the posterior distribution.\\
nfilt & & Number of filters used for photometric redshift fit.
\enddata
\tablecomments{All fluxes are corrected for the Galactic dust extinction, and set to magnitude zeropoint of $m_0=25$ i.e. $m=-2.5 \log({\rm flux})+m_0$. For sources with filters without data coverage (including those falling in the UVIS detector gap), the flux error arrays are set to $-99$.
}
\label{tab:col}
\end{deluxetable*}

\begin{longrotatetable}
\movetabledown=0.5in
\tabletypesize{\scriptsize} 
\tabcolsep=1.0pt 
\begin{deluxetable*}{ccccccccccccccccccccccc}
\tablecolumns{23} 
\tablewidth{0pt} 
\tablecaption{} 
\tablehead{
\colhead{field} &
\colhead{survey} &
\colhead{id} &
\colhead{ra} &
\colhead{dec} &
\colhead{x} &
\colhead{y} &
\colhead{kron\_radius} &
\colhead{a\_image} &
\colhead{b\_image} &
\colhead{theta} &
\colhead{class\_star} &
\colhead{flux\_radius} &
\colhead{flags} &
\colhead{f\_persist} &
\colhead{flag\_detect} &
\colhead{flux\_ratio0} &
\colhead{flux\_ratio1} &
\colhead{flux\_scale} &
\colhead{faper\_F160W} &
\colhead{eaper\_F160W} &
\colhead{faper\_F140W} &
\colhead{eaper\_F140W}
} 
\startdata
0624-6432& B17& 692& 95.86369& -64.52982& 1785.8214& 1078.0853& 3.500& 1.839& 1.717& 9.887& 0.976& 2.000& 0& 0& 1& 0.421& 0.563& 1.117& 4.752& 0.115& 4.792& 0.068\\
0624-6432& B17& 693& 95.871475& -64.529785& 1635.2957& 1079.6268& 6.707& 1.428& 1.261& -42.633& 0.379& 2.944& 0& 0& 1& 0.298& 0.440& 1.531& 0.767& 1.018& 0.628& 1.015\\
0624-6432& B17& 694& 95.876686& -64.52985& 1534.4542& 1077.1658& 5.278& 2.458& 1.239& -55.542& 0.425& 3.372& 0& 0& 1& 0.304& 0.490& 1.720& 0.969& 0.234& 0.643& 0.069\\
0624-6432& B17& 695& 95.88714& -64.529335& 1332.1267& 1100.2076& 3.994& 1.283& 1.026& -42.011& 0.452& 1.565& 0& 0& 1& 0.339& 0.707& 1.091& 0.489& 0.112& 0.382& 0.066\\
0624-6432& B17& 696& 95.85694& -64.529915& 1916.4974& 1073.7604& 5.134& 2.476& 1.288& -36.673& 0.437& 3.159& 0& 0& 1& 0.344& 0.389& 1.517& 1.084& 0.256& 0.884& 0.129\\
0624-6432& B17& 697& 95.88604& -64.52994& 1353.4473& 1073.0458& 4.331& 2.216& 1.712& -84.916& 0.560& 3.178& 0& 0& 1& 0.303& 0.397& 1.571& 1.612& 0.112& 1.545& 0.066\\
0624-6432& B17& 698& 95.91055& -64.52986& 879.05133& 1076.2301& 3.500& 0.932& 0.749& 88.795& 0.369& 1.728& 0& 0& 1& 0.306& 0.668& 1.065& 0.390& 0.248& 0.013& 0.067\\
0624-6432& B17& 699& 95.9241& -64.5299& 616.85004& 1074.6774& 3.500& 1.539& 1.437& 83.791& 0.119& 1.853& 0& 0& 1& 0.354& 0.594& 1.068& 1.630& 0.110& 1.706& 0.065\\
0624-6432& B17& 700& 95.86257& -64.52992& 1807.6199& 1073.5538& 3.500& 1.832& 1.494& -21.711& 0.923& 1.727& 0& 0& 1& 0.450& 0.626& 1.087& 2.620& 0.114& 3.050& 0.067\\
0624-6432& B17& 701& 95.872406& -64.52989& 1617.3138& 1075.0374& 5.676& 1.165& 0.947& -60.002& 0.414& 1.843& 0& 0& 1& 0.309& 0.584& 1.105& 0.432& 0.137& 0.010& 0.077\\
0624-6432& B17& 702& 95.88535& -64.52991& 1366.6926& 1074.1544& 6.775& 1.416& 0.989& 86.399& 0.351& 3.004& 0& 0& 1& 0.264& 0.385& 1.499& 0.532& 0.112& 0.418& 0.067\\
0624-6432& B17& 703& 95.87326& -64.52997& 1600.6573& 1071.4878& 3.500& 1.888& 1.687& -73.838& 0.982& 2.056& 0& 0& 1& 0.391& 0.555& 1.156& 2.944& 0.208& 2.969& 0.082\\
0624-6432& B17& 704& 95.8551& -64.529915& 1952.1763& 1073.8175& 6.801& 1.110& 0.775& 2.004& 0.368& 2.387& 0& 0& 1& 0.385& 0.549& 1.217& 0.490& 0.121& 0.464& 0.088\\
0624-6432& B17& 705& 95.93543& -64.52996& 397.5872& 1071.7129& 6.042& 1.946& 1.460& 5.871& 0.396& 3.622& 0& 0& 1& 0.287& 0.413& 1.840& 1.017& 0.145& 0.974& 0.101\\
0624-6432& B17& 706& 95.85896& -64.530045& 1877.3486& 1067.7806& 3.500& 1.772& 1.712& 76.289& 0.976& 1.847& 0& 0& 1& 0.406& 0.594& 1.096& 7.224& 0.113& 8.520& 0.068\\
0624-6432& B17& 707& 95.902954& -64.52934& 1026.032& 1099.8868& 7.204& 1.446& 0.778& 39.550& 0.348& 2.645& 0& 0& 1& 0.303& 0.373& 1.212& 0.488& 0.116& 0.348& 0.067\\
0624-6432& B17& 708& 95.87867& -64.530106& 1495.9816& 1065.265& 4.298& 1.891& 1.555& 57.190& 0.958& 2.140& 0& 0& 1& 0.341& 0.558& 1.196& 2.823& 0.114& 2.962& 0.067\\
0624-6432& B17& 709& 95.911705& -64.53009& 856.7665& 1066.0654& 3.500& 1.378& 1.289& -80.727& 0.013& 1.873& 0& 0& 1& 0.284& 0.569& 1.033& 1.037& 0.127& 0.910& 0.068\\
0624-6432& B17& 710& 95.90576& -64.53012& 971.83105& 1064.6478& 3.500& 1.329& 0.992& -48.639& 0.448& 1.772& 0& 0& 1& 0.422& 0.567& 1.004& 0.499& 0.112& 0.508& 0.066\\
0624-6432& B17& 711& 95.85113& -64.53013& 2029.095& 1064.1027& 3.918& 1.254& 1.156& -67.490& 0.352& 2.448& 0& 0& 1& 0.202& 0.371& 1.091& 0.588& 0.119& 0.210& 0.071\\
0624-6432& B17& 712& 95.89998& -64.53& 1083.6361& 1070.3765& 4.686& 1.537& 1.361& -70.912& 0.862& 2.002& 0& 0& 1& 0.280& 0.563& 1.129& 1.603& 0.110& 1.332& 0.066\\
0624-6432& B17& 713& 95.87182& -64.53016& 1628.6356& 1062.8197& 3.943& 1.475& 0.914& -69.848& 0.396& 1.830& 0& 0& 1& 0.330& 0.605& 1.111& 0.543& 0.114& -0.022& 0.068\\
0624-6432& B17& 714& 95.919334& -64.53017& 709.16614& 1062.4702& 5.654& 1.011& 0.949& -60.457& 0.548& 1.517& 0& 0& 1& 0.329& 0.666& 1.021& 0.398& 0.193& 0.353& 0.073\\
0624-6432& B17& 715& 95.88368& -64.53023& 1398.968& 1059.8787& 4.491& 1.063& 1.009& -4.259& 0.451& 1.908& 0& 0& 1& 0.261& 0.534& 1.000& 0.491& 0.112& 0.405& 0.087\\
0624-6432& B17& 716& 95.852745& -64.53028& 1997.6974& 1057.1389& 5.990& 1.244& 1.089& 36.919& 0.463& 2.739& 0& 0& 1& 0.348& 0.451& 1.393& 0.564& 0.116& 0.311& 0.069\\
0624-6432& B17& 717& 95.90474& -64.53029& 991.6068& 1057.0814& 6.129& 1.158& 1.015& -69.227& 0.359& 2.414& 2& 0& 1& 0.284& 0.452& 1.200& 0.498& 0.108& 0.400& 0.065\\
0624-6432& B17& 718& 95.87924& -64.53039& 1485.0106& 1052.5944& 6.526& 2.423& 1.591& 51.264& 0.516& 5.886& 0& 0& 1& 0.281& 0.405& 2.645& 1.341& 0.113& 1.187& 0.066\\
0624-6432& B17& 719& 95.910225& -64.530334& 885.33154& 1055.2196& 3.604& 1.306& 0.819& 60.341& 0.417& 1.865& 0& 0& 1& 0.379& 0.582& 1.063& 0.445& 0.112& 0.393& 0.066\\
0624-6432& B17& 720& 95.91241& -64.53033& 843.1588& 1055.3024& 5.104& 0.922& 0.676& -70.220& 0.425& 1.552& 0& 0& 1& 0.398& 0.730& 1.074& 0.330& 0.112& 0.013& 0.066\\
0624-6432& B17& 721& 95.92927& -64.53034& 516.8863& 1054.5538& 4.604& 1.266& 0.959& -33.048& 0.716& 1.867& 0& 0& 1& 0.325& 0.684& 1.230& 0.781& 0.381& 0.794& 0.189
\enddata
\tablecomments{Example 30 lines of the master catalog. The full table is available online.}
\label{tab:master}
\end{deluxetable*}

\end{longrotatetable}

\begin{longrotatetable}
\movetabledown=0.5in
\tabletypesize{\scriptsize} 
\tabcolsep=0.8pt 
\begin{deluxetable*}{cccccccccccccccccccccccccccc}
\tablecolumns{28} 
\tablewidth{0pt} 
\tablecaption{} 
\tablehead{
\colhead{field} &
\colhead{survey} &
\colhead{id} &
\colhead{f\_F300X} &
\colhead{e\_F300X} &
\colhead{f\_F350LP} &
\colhead{e\_F350LP} &
\colhead{f\_F435W} &
\colhead{e\_F435W} &
\colhead{...}  &
\colhead{f\_F125W} &
\colhead{e\_F125W} &
\colhead{f\_F140W} &
\colhead{e\_F140W} &
\colhead{f\_F160W} &
\colhead{e\_F160W} &
\colhead{f\_irac1} &
\colhead{e\_irac1} &
\colhead{f\_irac2} &
\colhead{e\_irac2} &
\colhead{f\_irac3} &
\colhead{e\_irac3} &
\colhead{f\_irac4} &
\colhead{e\_irac4} &
\colhead{flag\_irac1} &
\colhead{flag\_irac2} &
\colhead{flag\_irac3} &
\colhead{flag\_irac4}
} 
\startdata
1301+1300& B25& 1& 0.000& -99.000& 3.920& 1.272& 0.000& -99.000& ...& 19.766& 1.627& 22.364& 1.486& 25.674& 1.860& 0.000& -99.000& 0.000& -99.000& 0.000& -99.000& 0.000& -99.000& 0& 0& 0& 0\\
1301+1300& B25& 2& 0.000& -99.000& 1.017& 0.084& 0.000& -99.000& ...& 0.944& 5.237& 0.880& 10.597& 1.226& 4.983& 0.000& -99.000& 0.000& -99.000& 0.000& -99.000& 0.000& -99.000& 0& 0& 0& 0\\
1301+1300& B25& 3& 0.000& -99.000& 12.852& 0.108& 0.000& -99.000& ...& 75.957& 0.119& 90.714& 0.111& 106.165& 0.140& 0.000& -99.000& 0.000& -99.000& 0.000& -99.000& 0.000& -99.000& 0& 0& 0& 0\\
1301+1300& B25& 4& 0.000& -99.000& 1.068& 0.060& 0.000& -99.000& ...& 0.752& 25.984& 0.952& 24.338& 0.963& 26.682& 0.000& -99.000& 0.000& -99.000& 0.000& -99.000& 0.000& -99.000& 0& 0& 0& 0\\
1301+1300& B25& 5& 0.000& -99.000& 0.223& 0.067& 0.000& -99.000& ...& 0.587& 0.109& 0.476& 0.098& 0.499& 0.109& 24.253& 1.619& 0.000& -99.000& 0.000& -99.000& 0.000& -99.000& 0& 0& 0& 0\\
1301+1300& B25& 6& 0.000& -99.000& 0.162& 0.079& 0.000& -99.000& ...& 0.361& 8.465& 0.393& 10.226& 0.364& 9.321& 0.000& -99.000& 0.000& -99.000& 0.000& -99.000& 0.000& -99.000& 0& 0& 0& 0\\
1301+1300& B25& 7& 0.000& -99.000& 0.087& 0.072& 0.000& -99.000& ...& 0.650& 0.098& 0.697& 0.089& 0.802& 0.110& 0.000& -99.000& 0.000& -99.000& 0.000& -99.000& 0.000& -99.000& 0& 0& 0& 0\\
1301+1300& B25& 8& 0.000& -99.000& 9.506& 0.087& 0.000& -99.000& ...& 50.462& 0.096& 59.563& 0.088& 68.997& 0.112& 0.000& -99.000& 0.000& -99.000& 0.000& -99.000& 0.000& -99.000& 0& 0& 0& 0\\
1301+1300& B25& 9& 0.000& -99.000& 0.458& 0.280& 0.000& -99.000& ...& 1.503& 0.433& 1.900& 0.412& 1.818& 0.475& 0.000& -99.000& 0.000& -99.000& 0.000& -99.000& 0.000& -99.000& 0& 0& 0& 0\\
1301+1300& B25& 10& 0.000& -99.000& 0.526& 0.065& 0.000& -99.000& ...& 0.619& 0.068& 0.750& 0.063& 1.287& 0.080& 0.000& -99.000& 0.000& -99.000& 0.000& -99.000& 0.000& -99.000& 0& 0& 0& 0\\
1301+1300& B25& 11& 0.000& -99.000& 0.072& 0.088& 0.000& -99.000& ...& 0.117& 0.131& 0.217& 0.117& 0.210& 0.143& 0.000& -99.000& 0.000& -99.000& 0.000& -99.000& 0.000& -99.000& 0& 0& 0& 0\\
1301+1300& B25& 12& 0.000& -99.000& 0.555& 0.144& 0.000& -99.000& ...& 1.441& 0.260& 1.589& 0.492& 2.015& 0.240& 0.000& -99.000& 0.000& -99.000& 0.000& -99.000& 0.000& -99.000& 0& 0& 0& 0\\
1301+1300& B25& 13& 0.000& -99.000& 0.126& 0.041& 0.000& -99.000& ...& 0.114& 0.045& 0.097& 0.042& 0.034& 0.053& 0.000& -99.000& 0.000& -99.000& 0.000& -99.000& 0.000& -99.000& 0& 0& 0& 0\\
1301+1300& B25& 14& 0.000& -99.000& 0.167& 0.059& 0.000& -99.000& ...& 0.242& 0.072& 0.246& 0.066& 0.267& 0.082& 4.463& 1.659& 0.000& -99.000& 0.000& -99.000& 0.000& -99.000& 0& 0& 0& 0\\
1301+1300& B25& 15& 0.000& -99.000& 2.793& 0.124& 0.000& -99.000& ...& 6.383& 0.181& 6.958& 0.158& 7.284& 0.221& 0.000& -99.000& 0.000& -99.000& 0.000& -99.000& 0.000& -99.000& 0& 0& 0& 0\\
1301+1300& B25& 16& 0.000& -99.000& 0.666& 0.151& 0.000& -99.000& ...& 0.728& 0.160& 0.733& 0.148& 0.574& 0.188& 0.000& -99.000& 0.000& -99.000& 0.000& -99.000& 0.000& -99.000& 0& 0& 0& 0\\
1301+1300& B25& 17& 0.000& -99.000& 0.100& 0.102& 0.000& -99.000& ...& 0.140& 0.111& 0.329& 0.102& 0.167& 0.130& 0.000& -99.000& 0.000& -99.000& 0.000& -99.000& 0.000& -99.000& 0& 0& 0& 0\\
1301+1300& B25& 18& 0.000& -99.000& 0.104& 0.040& 0.000& -99.000& ...& 0.101& 0.044& 0.120& 0.041& 0.158& 0.052& 0.000& -99.000& 0.000& -99.000& 0.000& -99.000& 0.000& -99.000& 0& 0& 0& 0\\
1301+1300& B25& 19& 0.000& -99.000& 0.317& 0.050& 0.000& -99.000& ...& 0.418& 0.122& 0.454& 0.060& 0.448& 0.098& 0.000& -99.000& 0.000& -99.000& 0.000& -99.000& 0.000& -99.000& 0& 0& 0& 0\\
1301+1300& B25& 20& 0.000& -99.000& 0.101& 0.166& 0.000& -99.000& ...& 0.110& 0.227& 0.292& 0.206& 0.219& 0.255& 4.771& 1.622& 0.000& -99.000& 0.000& -99.000& 0.000& -99.000& 2& 0& 0& 0\\
1301+1300& B25& 21& 0.000& -99.000& 0.015& 0.041& 0.000& -99.000& ...& 0.116& 0.045& 0.104& 0.042& 0.057& 0.053& 0.000& -99.000& 0.000& -99.000& 0.000& -99.000& 0.000& -99.000& 0& 0& 0& 0\\
1301+1300& B25& 22& 0.000& -99.000& 3.828& 0.140& 0.000& -99.000& ...& 5.259& 0.188& 5.219& 0.172& 5.499& 0.212& 0.000& -99.000& 0.000& -99.000& 0.000& -99.000& 0.000& -99.000& 0& 0& 0& 0\\
1301+1300& B25& 23& 0.000& -99.000& -0.019& 0.040& 0.000& -99.000& ...& 0.122& 0.048& 0.112& 0.048& 0.078& 0.056& 0.000& -99.000& 0.000& -99.000& 0.000& -99.000& 0.000& -99.000& 0& 0& 0& 0\\
1301+1300& B25& 24& 0.000& -99.000& 0.143& 0.051& 0.000& -99.000& ...& 0.023& 0.057& 0.043& 0.053& 0.195& 0.067& 0.000& -99.000& 0.000& -99.000& 0.000& -99.000& 0.000& -99.000& 0& 0& 0& 0\\
1301+1300& B25& 25& 0.000& -99.000& 0.029& 0.057& 0.000& -99.000& ...& 0.166& 0.063& 0.264& 0.058& 0.185& 0.073& 0.000& -99.000& 0.000& -99.000& 0.000& -99.000& 0.000& -99.000& 0& 0& 0& 0\\
1301+1300& B25& 26& 0.000& -99.000& 28.192& 0.051& 0.000& -99.000& ...& 40.712& 0.053& 42.117& 0.049& 42.048& 0.063& 4.303& 1.621& 0.000& -99.000& 0.000& -99.000& 0.000& -99.000& 0& 0& 0& 0\\
1301+1300& B25& 27& 0.000& -99.000& 0.044& 0.040& 0.000& -99.000& ...& 0.030& 0.045& 0.109& 0.041& 0.107& 0.053& 0.000& -99.000& 0.000& -99.000& 0.000& -99.000& 0.000& -99.000& 0& 0& 0& 0\\
1301+1300& B25& 28& 0.000& -99.000& -0.010& 0.045& 0.000& -99.000& ...& -0.011& 0.057& 0.070& 0.052& 0.111& 0.065& 8.442& 1.684& 0.000& -99.000& 0.000& -99.000& 0.000& -99.000& 0& 0& 0& 0\\
1301+1300& B25& 29& 0.000& -99.000& 0.173& 0.055& 0.000& -99.000& ...& 0.186& 0.058& 0.184& 0.054& 0.209& 0.069& 0.000& -99.000& 0.000& -99.000& 0.000& -99.000& 0.000& -99.000& 0& 0& 0& 0\\
1301+1300& B25& 30& 0.000& -99.000& 0.151& 0.060& 0.000& -99.000& ...& 0.117& 0.113& 0.112& 0.180& 0.215& 0.107& 11.230& 1.649& 0.000& -99.000& 0.000& -99.000& 0.000& -99.000& 0& 0& 0& 0
\enddata
\tablecomments{Example 30 lines of the photometric flux catalog. Middle columns are omitted to be fitted in the manuscript size. The full table is available online.}
\label{tab:phot}
\end{deluxetable*}
\end{longrotatetable}

\begin{longrotatetable}
\movetabledown=0.0in
\tabletypesize{\scriptsize} 
\tabcolsep=4pt
\begin{deluxetable*}{cccccccccccccccccccccccc}
\tablecolumns{24} 
\tablewidth{0pt} 
\tablecaption{} 
\tablehead{
\colhead{field} &
\colhead{survey} &
\colhead{id} &
\colhead{zspec} &
\colhead{flag\_zspec} &
\colhead{z16} &
\colhead{z50} &
\colhead{z84} &
\colhead{zpeak} &
\colhead{MUV16} &
\colhead{MUV50} &
\colhead{MUV84} &
\colhead{MUVpeak} &
\colhead{zset} &
\colhead{plow} &
\colhead{phigh} &
\colhead{UVbeta\_lambda} &
\colhead{U-V} &
\colhead{B-V} &
\colhead{V-J} &
\colhead{z-J} &
\colhead{chi2} &
\colhead{zmc} &
\colhead{nfilt}
} 
\startdata
1301+1300& B25& 57& -99.0& -99.0& 0.45& 1.6& 2.59& 1.42& -14.014& -18.051& -19.691& -17.651& 6.5& 1.0& 0.0& -2.278& -0.267& 0.127& -0.631& -0.078& 0.701& 4.137& 7\\
1301+1300& B25& 58& -99.0& -99.0& 0.36& 2.28& 2.39& 2.29& -12.952& -18.837& -18.999& -18.852& 6.5& 1.0& 0.0& -2.101& -0.012& 0.132& -0.24& -0.004& 2.783& 2.354& 7\\
1301+1300& B25& 59& -99.0& -99.0& 1.27& 1.64& 3.41& 1.39& -16.541& -17.397& -19.9& -16.842& 6.5& 0.911& 0.089& -2.556& -0.444& 0.263& -1.311& -0.167& 5.349& 1.349& 7\\
1301+1300& B25& 60& -99.0& -99.0& 0.9& 3.94& 9.97& 0.01& -16.984& -21.969& -25.185& -5.622& 6.5& 0.651& 0.349& -2.451& 1.306& 0.548& -1.644& -0.167& 2.997& 0.497& 7\\
1301+1300& B25& 61& -99.0& -99.0& 0.84& 1.14& 1.38& 1.27& -17.346& -18.337& -18.97& -18.694& 6.5& 1.0& 0.0& -1.382& 1.071& 0.495& 1.187& 0.434& 0.087& 1.151& 7\\
1301+1300& B25& 62& -99.0& -99.0& 0.38& 1.35& 3.86& 0.39& -14.587& -18.545& -22.13& -14.661& 6.5& 1.0& 0.0& -2.301& -0.195& 0.214& -0.333& 0.022& 2.982& 1.461& 7\\
1301+1300& B25& 63& -99.0& -99.0& 0.89& 2.59& 7.0& 1.34& -14.368& -17.942& -21.38& -15.709& 6.5& 0.81& 0.19& -2.551& -0.054& 0.462& 0.167& 0.18& 1.777& 0.368& 7\\
1301+1300& B25& 64& -99.0& -99.0& 0.88& 1.68& 2.65& 1.44& -16.78& -18.915& -20.469& -18.397& 6.5& 0.986& 0.014& -2.555& -0.459& 0.252& -1.299& -0.167& 1.277& 1.659& 7\\
1301+1300& B25& 65& -99.0& -99.0& 1.32& 5.28& 7.91& 1.39& -17.235& -21.979& -23.38& -17.407& 6.5& 0.614& 0.385& -2.556& -0.444& 0.263& -1.309& -0.166& 5.218& 5.612& 7\\
1301+1300& B25& 66& -99.0& -99.0& 0.75& 3.4& 7.46& 0.18& -8.875& -13.931& -16.652& -4.797& 6.5& 0.692& 0.307& -0.517& 2.287& 1.083& 2.204& 0.694& 0.652& 2.404& 7\\
1301+1300& B25& 67& -99.0& -99.0& 0.08& 0.27& 2.78& 2.78& -9.981& -13.057& -20.423& -20.423& 6.5& 1.0& 0.0& -2.553& -0.456& 0.253& -1.295& -0.166& 3.364& 0.239& 7\\
1301+1300& B25& 68& -99.0& -99.0& 0.33& 2.33& 3.22& 2.78& -12.639& -18.844& -19.956& -19.45& 6.5& 1.0& 0.0& -2.555& -0.457& 0.253& -1.3& -0.166& 1.536& 3.525& 7\\
1301+1300& B25& 69& -99.0& -99.0& 0.77& 2.62& 4.45& 0.4& -15.308& -19.383& -21.211& -13.324& 6.5& 0.97& 0.03& -2.5& -0.265& 0.316& 0.35& 0.438& 0.63& 3.962& 7\\
1301+1300& B25& 70& -99.0& -99.0& 0.4& 1.64& 3.71& 3.57& -14.925& -19.389& -22.183& -22.05& 6.5& 1.0& 0.0& -2.32& -0.122& 0.275& -0.085& 0.169& 1.089& 2.971& 7\\
1301+1300& B25& 71& -99.0& -99.0& 1.35& 4.58& 9.2& 0.39& -11.146& -15.322& -17.739& -7.262& 6.5& 0.653& 0.346& 0.9& 1.263& 0.676& 1.872& 0.666& 4.044& 5.127& 7\\
1301+1300& B25& 72& -99.0& -99.0& 1.05& 3.48& 5.97& 1.69& -14.603& -18.659& -20.528& -16.187& 6.5& 0.89& 0.109& -2.553& -0.422& 0.276& -1.322& -0.166& 1.91& 4.119& 7\\
1301+1300& B25& 73& -99.0& -99.0& 2.54& 10.05& 11.35& 10.74& -15.673& -20.43& -20.851& -20.66& 6.5& 0.272& 0.728& -2.553& -0.436& 0.266& -1.308& -0.165& 1.235& 10.784& 7\\
1301+1300& B25& 74& -99.0& -99.0& 2.4& 2.91& 3.13& 2.82& -20.06& -20.722& -20.973& -20.614& 6.5& 1.0& 0.0& -2.554& -0.468& 0.245& -1.289& -0.166& 2.441& 3.214& 7\\
1301+1300& B25& 75& -99.0& -99.0& 2.2& 9.64& 10.99& 10.51& -15.715& -20.82& -21.273& -21.119& 6.5& 0.317& 0.682& -2.555& -0.44& 0.265& -1.312& -0.166& 4.164& 2.412& 7\\
1301+1300& B25& 76& -99.0& -99.0& 0.68& 1.31& 3.52& 1.14& -17.405& -19.522& -22.888& -19.063& 6.5& 1.0& 0.0& -1.946& 0.335& 0.114& 0.527& 0.248& 0.198& 3.631& 7\\
1301+1300& B25& 77& -99.0& -99.0& 0.08& 0.29& 1.44& 0.01& -10.652& -13.924& -18.857& -5.95& 6.5& 1.0& 0.0& -2.346& 1.487& 0.655& 0.79& 0.672& 7.415& 1.832& 7\\
1301+1300& B25& 78& -99.0& -99.0& 0.31& 2.29& 3.42& 3.52& -12.636& -18.956& -20.335& -20.435& 6.5& 1.0& 0.0& -2.555& -0.451& 0.257& -1.304& -0.166& 2.371& 2.553& 7\\
1301+1300& B25& 79& -99.0& -99.0& 0.67& 2.59& 4.2& 1.15& -14.723& -19.195& -20.863& -16.455& 6.5& 0.994& 0.006& -2.124& -0.098& 0.061& -0.338& -0.048& 0.375& 3.022& 7\\
1301+1300& B25& 80& -99.0& -99.0& 0.05& 0.15& 0.7& 0.12& 0.157& -2.477& -6.804& -1.921& 6.5& 1.0& 0.0& 8.633& 2.27& 1.024& 1.211& 0.265& 3.927& 0.107& 7\\
1301+1300& B25& 81& -99.0& -99.0& 1.24& 4.66& 7.07& 6.21& -13.482& -18.0& -19.445& -18.995& 6.5& 0.758& 0.241& -1.961& 0.166& 0.002& -0.107& -0.031& 0.139& 2.506& 7\\
1301+1300& B25& 82& -99.0& -99.0& 0.35& 2.29& 2.35& 2.35& -14.934& -20.914& -21.002& -21.002& 6.5& 1.0& 0.0& -2.184& -0.071& 0.154& 0.116& 0.231& 2.958& 2.335& 7\\
1301+1300& B25& 83& -99.0& -99.0& 1.95& 8.3& 9.96& 9.64& -14.769& -19.767& -20.398& -20.285& 6.5& 0.363& 0.637& -2.552& -0.457& 0.251& -1.291& -0.165& 0.293& 8.409& 7\\
1301+1300& B25& 84& -99.0& -99.0& 0.5& 0.82& 1.16& 0.64& -17.134& -18.661& -19.786& -17.884& 6.5& 1.0& 0.0& -1.977& 0.779& 0.574& 0.701& 0.085& 5.098& 1.232& 7\\
1301+1300& B25& 85& -99.0& -99.0& 1.25& 6.26& 8.2& 7.89& -14.113& -19.628& -20.564& -20.43& 6.5& 0.513& 0.486& -2.555& -0.448& 0.26& -1.307& -0.166& 1.326& 7.87& 7\\
1301+1300& B25& 86& -99.0& -99.0& 0.4& 0.51& 3.57& 0.45& -18.165& -18.879& -25.29& -18.508& 6.5& 1.0& 0.0& -2.511& -0.272& 0.333& 0.08& 0.4& 2.068& 0.43& 7
\enddata
\tablecomments{Example 30 lines of the photometric redshift catalog. The full table is available online.}
\label{tab:eazy}
\end{deluxetable*}

\end{longrotatetable}


\bibliographystyle{apj}
\bibliography{./adssample}

\end{document}